\documentclass[11pt]{amsart}

\usepackage{amssymb,amsmath}
\usepackage{latexsym}
\usepackage{graphicx}
\usepackage{geometry}  
\usepackage{caption}
\usepackage{subcaption}
\usepackage{color}

\newcommand{\ket}[1]{|\kern.3ex#1\kern.3ex\rangle}
\newcommand{\bra}[1]{\langle\kern.3ex #1 \kern.3ex|}
\newcommand{\scalar}[2]{\langle\kern.3ex #1 \kern.3ex|\kern.3ex#2\kern.3ex\rangle}
\newcommand{\norm}[1]{\|\kern.3ex#1\kern.3ex \|}

\def\lg{\langle }
\def\rg{\rangle }

\def\ud{\mathrm{d}}

\begin{document}
\title[]{Clocks and dynamics\\ in quantum models of gravity}

\author{Przemys\l aw Ma\l kiewicz}

\address{APC, UMR 7164 CNRS, Univ Paris  Diderot, Sorbonne Paris Cit\'e, 75205 Paris, France}\email{Przemyslaw.Malkiewicz@apc.univ-paris7.fr}

\address{National Centre for Nuclear Research, Ho\.za 69, 00-681 Warsaw, Poland} \email{Przemyslaw.Malkiewicz@fuw.edu.pl}

\date{\today}
\begin{abstract}
We investigate how the quantum dynamics of gravitational models depends on the internal clock employed in quantisation. Our previous result on the quantum Friedmann-Lemaitre model \cite{2M15} demonstrates that almost all physical predictions from the quantum cosmological dynamics, e.g. the scale of the quantum bounce or the number of bounces, depend on the choice of internal clock. In the present paper we show that there exists an important prediction concerning the quantum dynamics which holds in all internal clocks. Namely, we prove that any quantum state asymptotically, i.e. away from a quantum interaction regime, admits a unique classical limit described by unique classical solutions irrespectively of the internal clock used for dynamical description of the given state. We demonstrate this property explicitly for the semiclassical Bianchi Type I model, which includes the semiclassical Friedmann-Lemaitre model of \cite{2M15} as a special case. Our methods include the theory of pseudo-canonical transformations and the phase space portraits based on coherent states. As a by-product of the pursuit of our main goal, we elaborate the semiclassical description of anisotropic singularity resolution.
\end{abstract}
\maketitle

\section{Introduction}
Canonical formalism of general relativity admits the time re-parametrisation invariance which is established by means of a Hamiltonian constraint. Time is no longer a fundamental entity but rather an auxiliary parameter without physical significance. The physical measure of evolution is provided by ambiguously chosen physical degrees of freedom, the so-called internal clocks. The concept of internal clock is not problematic at the classical level as a unique classical evolution can be expressed in terms of many different clocks. However, quantum mechanics seems to rely crucially on a fixed external time and it is not at all obvious how the replacement of this fundamental entity by internal clocks affects the nature of quantum evolution. 

In the present paper we study the concept of quantum dynamics with respect to internal clocks in the context of a finite-dimensional model of general relativity. The Hamiltonian constraint formalism is brought to the unconstrained Hamiltonian formalism by making a choice of internal clock and reducing the number of canonical variables (see Sec II for more details). The reduced formalism is next quantised. This approach is called the reduced phase space approach. Although we use this particular approach, we expect that the obtained results should be universal and at least qualitatively approach-independent. In particular we do not expect that different results can be derived from the Dirac approach. The latter consists of first quantisation of the Hamiltonian constraint and then solving the respective quantum constraint equation. Solutions to the quantum constraint equation should be equipped with a new inner product, which can be shown (at least in known cases) to be identical with a choice of internal clock at the quantum level \cite{Ku}. Let us add that the concept of ``inner dynamics" is also discussed within the Feynman quantisation, see e.g. an early treatment by Misner \cite{misner0}.

Our goal is to look for possible physical dissimilarities between quantum dynamics expressed in different internal clocks. Some of those dissimilarities can be made apparent only by means of an involved and fully quantum analysis (see \cite{1M15,M12}). Others can be found by means of a semiclassical analysis and we focus on the latter. We investigate the quantum dynamics of the Bianchi Type I universe filled with perfect fluid. The quantum dynamics in a fixed internal clock was derived in \cite{BI} and for some details the reader is referred to the original paper, nevertheless we try to keep the presentation self-contained. We examine semiclassical features of the quantum dynamics by means of the phase space portraits based on coherent states. The theory of so-called pseudo-canonical transformations \cite{1M15} plays a key role in switching the quantum dynamics to other internal clocks and making the comparison with the original dynamics.

The result of the present paper extends, in a very important way, our previous result \cite{2M15} which concerns the effect of the choice of internal clock on the quantum dynamics of the Friedmann-Lemaitre universe which is a special case of the Bianchi Type I universe. Therein, we found that the quantum dynamics largely depends on the chosen internal clock. The feature that exists in all internal clocks, is a quantum bounce which replaces the classical singularity. As a further matter, we found that away from the bounce the semiclassical solutions admit a classical limit which is independent of the choice of internal clock. The latter can be seen as a necessary property of semiclassical dynamics in two-dimensional phase spaces because in this case the contour plot of the semiclassical Hamiltonian that away from the bounce becomes classical, defines the asymptotic solutions unambiguously. This is not the case for higher-dimensional phase spaces, where infinitely many solutions may correspond to a given contour of the Hamiltonian. In the present work, we study a higher dimensional model in order to find out whether the described property holds in higher dimensions too. The issue is whether some form of determinism is present in quantum models of gravity devoid of a preferred internal clock.

The paper is organised as follows. In Sec II we introduce the concepts of internal clock and reduced phase space in canonical relativity. Then we apply these concepts to define the reduced canonical formalism for the Bianchi Type I model. In Sec III we discuss the theory of pseudo-canonical transformations. A special attention is devoted to a subclass of the so-called special pseudo-canonical transformations. Next we apply them to the Bianchi I model. In Sec IV we show quantisation of the model and discuss some quantum level differences induced by different choices of internal clock. The discussion is based on the special pseudo-canonical transformations. In Sec V we describe the method of phase space portraits and use it to plot the semiclassical dynamics of the examined model. Then we explain how the method can be used for making the comparison between semiclassical dynamics in different internal clocks. Sec VI includes the main result of the paper. We discuss several examples of internal clocks and compare the associated semiclassical dynamics. In Sec VII we generalise the invariant property of the portraits examined in Sec VI to all models and to all internal clocks. We conclude in Sec VIII. 

\section{Clocks and dynamics in Hamiltonian constraint formalism}
In this section we first make general remarks concerning internal clocks in canonical relativity and then discuss a specific model which is further studied in the next sections. We assume that the spacetime manifold $\mathcal{M}$ can be foliated, $\mathcal{M}=\Sigma\times\mathbb{R}$, where $\Sigma$ is a space-like leaf and $\mathbb{R}$ is a time manifold. We follow the canonical formalism of general relativity by Arnowitt, Deser and Misner \cite{adm}, which involves a Hamiltonian constraint,
\begin{equation}\label{gr}
\mathrm{H}=\int_{\Sigma} NC^0+\beta_iC^i,~~~\ C^{\mu}[g_{ab},p^{cd}](x)\approx 0,
\end{equation}
which is a sum of first-class constraints, $C^{0}$ and $C^{i}$. For brevity, we omit the optional matter part of the Hamiltonian. The three-metric $g_{ab}(x)$ is induced on $\Sigma$ from the metric on $\mathcal{M}$, its conjugate momentum $p^{ab}(x)$ encodes the extrinsic curvature of $\Sigma$ in $\mathcal{M}$. The constraints $C^{0}$ and $C^{i}$ are a density-weighted scalar and a density-weighted three-vector in $\Sigma$, respectively. The lapse function $N$ and the shift vector $\beta_i$ are the Lagrange multipliers which enforce the constraint equations, $C^{\mu}=0$. Since the described formalism involves constraints, it must include redundant variables. 

\subsection{General relativity as parametrised field theory}
The so called paradigm of general relativity as a parametrised field theory \cite{Ku} helps to understand the various roles of the variables in Eq. (\ref{gr}). It assumes that only some variables describe physical degrees of freedom while others play a role of internal space-time coordinates or are simply redundant and can be removed. The division of roles is made by means of a canonical transformation:
\begin{equation}\label{can}
(g_{ab},p^{cd})~\longrightarrow~(X_{\mu},P^{\mu},\phi_r,\pi^r)
\end{equation}
which introduces new canonical variables such that the constraints may be linearised with respect to $P^{\mu}$'s and read:
\begin{equation}\label{con}
C^{\mu}\approx P^{\mu}+h^{\mu}(X_{\nu},\phi_r,\pi^r)
\end{equation}
(where ``$\approx$" means ``equal up to a non-vanishing factor" which is a function of phase space variables). Now, $X_0$ and $X_i$'s may be identified with the internal time and space coordinates respectively, while $\phi_r$'s and $\pi^r$'s are interpreted as the physical degrees of freedom. $P^{\mu}$'s is removed from the description as a redundant piece of information. Indeed, if one sets
\begin{equation}
X_0(t,x^i):=t,~~ X_i(t,x^i):=x^i
\end{equation}
the evolution equations for $\phi_r(t,x^i)$'s and $\pi^r(t,x^i)$'s read:
\begin{equation}
\frac{\ud}{\ud t}\phi_r=\{\phi_r, \mathrm{H}_{true}\}_{red},~~\frac{\ud}{\ud t}\pi_r=\{\pi_r, \mathrm{H}_{true}\}_{red}
\end{equation}
where the non-vanishing Hamiltonian reads\begin{equation}\label{true}
\mathrm{H}_{true}:=\int_{\Sigma} h^{0} (t, x^{i},\phi_r,\pi^r)~\ud^3x,
\end{equation}
and the reduced Poisson bracket reads:
\begin{equation}\label{true1}
\{\phi_r,\pi^s\}_{red}=\delta_r^{~s}
\end{equation}
The space $(\phi_r,\pi^r)$ is called the reduced phase space. Note that $h^{0}$ is a density-weighted scalar and $\mathrm{H}_{true}$ is a scalar. 

We notice that the canonical variables $(X_{\mu},P^{\mu},\phi_r,\pi^r)$ are not the only ones which fulfil the condition (\ref{con}) and let us denote another set of variables satisfying (\ref{con}) by  $(\tilde{X}_{\mu},\tilde{P}^{\mu},\tilde{\phi}_r,\tilde{\pi}^r)$. By the virtue of construction the respective Hamiltonian $\tilde{\mathrm{H}}_{true}$ analogous to (\ref{true}) generates the same evolution of the system, which can be verified after having solved the respective Hamilton equations by means of the coordinate relation between $(\tilde{X}_{\mu},\tilde{\phi}_r,\tilde{\pi}^r)$ and  $(X_{\mu},\phi_r,\pi^r)$. 

We shall call the internal time coordinate $X_0$ ``internal clock". Its choice is largely ambiguous. The concept of internal clock is much richer than the concept of time coordinate in a fixed spacetime. Indeed, the reduced phase space $(\phi_r,\pi^r)$ describes many spatial geometries which are admissible for a fixed value of internal clock. In other words, an internal clock is not restricted to a single spacetime and provides a unified ``time coordinate" for all admissible spacetimes. It is further contrasted with the time coordinate by the property that the value of clock $X_0$ carries a physical meaning and is needed for the full reconstruction of the geometry of the system. 

\subsection{Bianchi type I model}
Let us consider a spatially homogeneous universe with vanishing spatial curvature and of toroidal spatial topology, $\mathcal{M}=\mathbb{T}^3\times\mathbb{R}$. The line element reads:
\begin{equation}
\ud s^2=-N^2\ud t^2+a_1^2(\ud x^1)^2+a_2^2(\ud x^2)^2+a_3^2(\ud x^3)^2
\end{equation}
where $\int_{\mathbb{S}} dx^i=1$ with the integration along the closed curve given by $x^{j}=const,~\forall j\neq i$. The universe is filed with a perfect fluid satisfying the linear equation of state $\mathrm{p}=w\mathrm{\rho},~w<1$. For the description of the fluid we use the Schutz formalism \cite{Sch}. It can be shown that the Hamiltonian constraint reads  \cite{BI}:
\begin{align}\label{cons0}
\mathrm{H}=NC^0,~~~C^0=\frac{c_w^{\frac{1-w}{2}}q^{\frac{1-w}{2}}}{24}\left(24q^2p_T-c_w^2q^2p^2+p_+^2+p_-^2\right),
\end{align}
where the geometrical variables read
\begin{align}\nonumber
q=(a_1a_2a_3)^{\frac{1-w}{2}}>0,~~~~p=\frac{8}{3(1-w)}(a_1a_2a_3)^{\frac{w-1}{2}}\frac{\dot{(a_1a_2a_3)}}{3N},\\ \label{int1}
p_+=-\frac{2}{N}a_1\dot{(a_2a_3)} -\frac{2}{N}a_2\dot{(a_3a_1)} +\frac{4}{N}a_3\dot{(a_1a_2)},~\beta_+=\frac{1}{6}\ln \frac{a_1a_2}{a_3^2},\\ \nonumber p_-=-\frac{2\sqrt{3}}{N}a_1\dot{(a_2a_3)} +\frac{2\sqrt{3}}{N}a_2\dot{(a_3a_1)},~~\beta_-=\frac{1}{2\sqrt{3}}\ln\frac{a_1}{a_2} .
\end{align}
The variables $q>0$ and $p$ describe isotropic evolution, whereas the variables $p_+,\beta_+$ and $p_-,\beta_-$, proposed by Misner (see e.g. \cite{cwm3}), describe anisotropic evolution of the space-like surface. The momentum $p_T$ and the canonically conjugate $T$ are fluid variables. The constant $c_w=\frac{3(1-w)}{2}$ parametrises the types of fluid, e.g. for radiation $c_w=1$.

\subsubsection{De-parameterisation}
We notice that the constraint of Eq. (\ref{cons0}) is already linear in the momentum $p_T$ and can be given the form of Eq. (\ref{con}),
\begin{align}
C^0=c_w^{\frac{1-w}{2}}q^{\frac{5-w}{2}}\left(p_T-\frac{c_w^2}{24}p^2+\frac{p_+^2+p_-^2}{24 q^2}\right)\approx p_T-\frac{c_w^2}{24}p^2+\frac{p_+^2+p_-^2}{24 q^2}~.
\end{align}
We choose $T$ for the internal clock and $(q,p,\beta_{\pm},p_{\pm})$ for the physical variables. The momentum $p_T$ is removed from the formalism. Hence in analogy to Eq. (\ref{true}) the true Hamiltonian reads \cite{BI}
\begin{equation}\label{ham0}
\mathrm{H}_{true}=\frac{c_w^2}{24}p^2-\frac{1}{24}\frac{p_+^2+p_-^2}{q^2}.
\end{equation}
(where for convenience we choose `$-T$' rather than `$T$' for the clock). The true Hamiltonian (\ref{ham0}) generates dynamics in the six-dimensional reduced phase space $(q,p,\beta_{\pm},p_{\pm})\in\mathbb{R}_+\times\mathbb{R}^5$ with respect to the internal clock that is associated with the fluid. 

\subsubsection{Dynamics}
The dynamics generated by (\ref{ham0}) exhibits a big bang singularity. The overall contraction of space, whose kinetic energy is described by the term $p^2$, is fuelled by the energy of anisotropic evolution $\frac{p_+^2+p_-^2}{q^2}$, which is growing unboundedly as $q$ decreases. At a finite value of the clock (and within a finite proper time) $q\rightarrow 0$ and $p\rightarrow \pm\infty$. The model is explicitly integrable and the isotropic part of evolution reads:
\begin{align}\label{sol0}
p(T)=\frac{2\mathrm{H}_{true}(T-T_0)}{\sqrt{\frac{c_w^2 \mathrm{H}_{true}}{6}(T-T_0)^2-\frac{k^2}{24\mathrm{H}_{true}}}},~~q(T)=\sqrt{\frac{c_w^2 \mathrm{H}_{true}}{6}(T-T_0)^2-\frac{k^2}{24\mathrm{H}_{true}}}
\end{align}
where $k^2=p_+^2+p_-^2$. For describing the anisotropic part of evolution, new canonical anisotropic variables, $(k, p_k,\alpha, p_{\alpha})\in\mathbb{R}_+\times\mathbb{R}\times[0,2\pi)\times\mathbb{R}$, are useful. They read:
\begin{align}
k\cos\alpha=p_+,~~k\sin\alpha=p_-,~~ p_k=-\frac{1}{k}\left(\beta_+p_++\beta_-p_-\right),~~p_{\alpha}=\beta_+p_--\beta_-p_+
\end{align}
The Hamiltonian (\ref{ham0}) reads now
\begin{equation}\label{ham00}
\mathrm{H}_{true}=\frac{c_w^2}{24}p^2-\frac{1}{24}\frac{k^2}{q^2}.
\end{equation}
The anisotropic variables $k$, $\alpha$ and $p_{\alpha}$ are constant, whereas $p_k$ reads:
\begin{align}\label{shape}
p_k(T)=\frac{1}{2c_w}\ln\left[\frac{(T-T_0)-\frac{k}{2c_w\mathrm{H}_{true}}}{(T-T_0)+\frac{k}{2c_w\mathrm{H}_{true}}}\right]+const.
\end{align}
We shall call `$p_k$' the {\it shape function}. It parametrises the ratios between the three scale factors $\frac{a_1}{a_2},\frac{a_2}{a_3},\frac{a_1}{a_3}$ and is independent of the value of the mean scale factor (see Appendix \ref{A} for specific formulas). We notice that in the isotropy limit $k\to 0$, the Hamiltonian (\ref{ham00}) describes the Friedmann-Lemaitre model (or, a free particle on a half-line) studied in \cite{2M15} and the shape function (\ref{shape}) becomes constant in time. The shape function is the only dynamical variable by which the set of dynamical variables, $q$ and $p$, studied already in \cite{2M15} is enlarged in the present model. Its quantum dynamics will be examined with respect to various internal clocks. 

\begin{figure}[t]
\begin{tabular}{cc}
\includegraphics[width=0.4\textwidth]{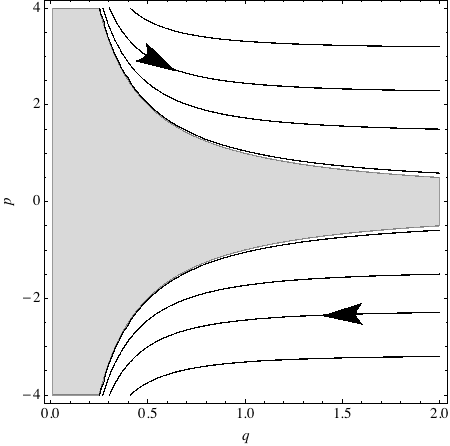}
\hspace{1cm}
\includegraphics[width=0.4\textwidth]{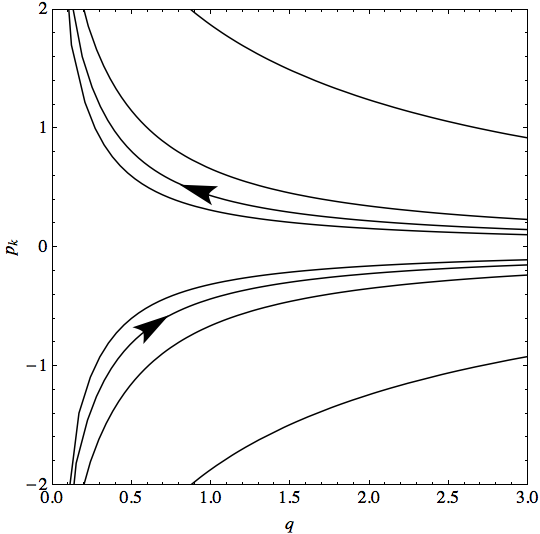}

\end{tabular}
\caption{\small Phase space portrait of the classical dynamics in the planes $(q,p)$ and $(q,p_k)$ for $k=1$, $c_w=1$ and $\mathrm{H}_{true}=\frac{1}{240},\frac{1}{12},\frac{5}{24}, \frac{5}{12}$. The classical dynamics does not occur for $\mathrm{H}_{true}<0$ in the shaded region.} 
\label{figure1}
\end{figure}

We note that the dynamics generated by (\ref{ham00}) is constrained by the {\it positivity constraint}, $$\mathrm{H}_{true}>0.$$ The phase space region $\mathrm{H}_{true}<0$ is unphysical and corresponds to the negative energy of the fluid. It does not matter how one defines the Hamiltonian inside the unphysical region. The existence of the unphysical region is a topological property of the phase space, i.e. it is independent of the coordinate system. It is caused by the shear-fuelled strongly singular dynamics of the anisotropic model. It makes the solutions divided into two disconnected branches in the phase space, an expanding and contracting one. This topological property is depicted in Fig. \ref{figure1}. The positivity constraint has to be included at the quantum level. As we shall shortly see, the unphysical region provides a set of geometrical configurations of the universe, through which quantisation can establish a connection between the two branches.

\section{Theory of pseudo-canonical transformations}\label{gentheory}
The theory of pseudo-canonical transformations was previously discussed in \cite{1M15,2M15}. Below we recall the definition of pseudo-canonical transformations and discuss them with emphasis on their group structure. We construct a simple method for formulating the dynamics of a single integrable system with respect to many clocks. Then we apply the method to the Bianchi type I model which was de-parametrised with respect to the fluid variable $T$ in the previous section.

\subsection{Pseudo-canonical transformations}
The theory of contact (or, time-dependent canonical) transformations \cite{AbMa} is based in the contact manifold, which is a product of phase space $\mathbb{P}$ and time manifold $\mathbb{R}$, $\mathcal{C}=\mathbb{R}\times\mathbb{P}$. Contact coordinates include the time coordinate `$t$' and the canonical coordinates `$(q^I,p_I)$'. The contact manifold is equipped with a contact form $\omega_{\mathcal{C}}$ which takes the following form
\begin{equation}\label{conf0}
\omega_{\mathcal{C}}=\ud q^I\ud p_I-\ud t\ud \mathrm{h}\end{equation}
where $\mathrm{h}$ is the Hamiltonian. The contact form encodes Hamilton's equations of motion in time $t$, for canonical coordinates $q^I$ and $p_I$ and generated by the Hamiltonian $\mathrm{h}$,
\begin{equation}\label{hameq}
\frac{\ud q^I}{\ud t}=\frac{\partial \mathrm{h}}{\partial p_I},~~\frac{\ud p^I}{\ud t}=-\frac{\partial \mathrm{h}}{\partial q^I}.
\end{equation}
Any solution $t\mapsto (t,q^I,p_I)$ to eqs (\ref{hameq}) defines a curve in $\mathcal{C}$ whose tangent vector is null with respect to $\omega_{\mathcal{C}}$ at any point.

Contact transformations are coordinate transformations in $\mathcal{C}$, which preserve the time coordinate $t$ and the canonical form of $\omega_{\mathcal{C}}$:
\begin{equation}\label{conf1}
(q^I,p_I,t)\mapsto (\bar{q}^I,\bar{p}_I,t):~~\omega_{\mathcal{C}}=\ud \bar{q}^I\ud  \bar{p}_I-\ud t\ud  \bar{\mathrm{h}},
\end{equation}
where $\bar{q}^I ~,\bar{p}_I$ are new canonical variables which depend on $q^I,~ p_I$ and $t$ and whose dynamics is generated by the new Hamiltonian $\bar{\mathrm{h}}$. Note that the time coordinate $t$ is preserved.

The idea of pseudo-canonical transformations (PTs), which was first considered in \cite{ADM61}, is to extend the contact transformations to allow for time coordinate transformations as well. The only requirement is that the canonical form of $\omega_{\mathcal{C}}$ is preserved, namely:
\begin{equation}\label{conf2}
(q^I,p_I,t)\mapsto (\bar{q}^I,\bar{p}_I,\bar{t}):~~\omega_{\mathcal{C}}=\ud \bar{q}^I\ud  \bar{p}_I-\ud \bar{t}\ud  \bar{\mathrm{h}},
\end{equation}
where $\bar{q}^I ~,\bar{p}_I$ are new canonical variables, $\bar{t}$ is a new time coordinate and $\bar{\mathrm{h}}$ is a new Hamiltonian. All new coordinates are functions of $q^I,~ p_I,~t$. The respective Hamilton equations read
\begin{equation}
\frac{\ud \bar{q}^I}{\ud \bar{t}}=\frac{\partial \bar{\mathrm{h}}}{\partial \bar{p}_I},~~\frac{\ud \bar{p}^I}{\ud \bar{t}}=-\frac{\partial \bar{\mathrm{h}}}{\partial \bar{q}^I}.
\end{equation}
Up to time re-parametrisation, they generate exactly the same curves in $\mathcal{C}$ as eqs (\ref{hameq}). Physically, it means that the motion of the system is exactly the same as before. Technically, the canonical structure is induced by the choice of time coordinate and the coordinates $q^I$ and $p_I$ are canonical with respect to $t$ but, in general, they are not canonical with respect to $\bar{t}$ (and vice versa, the coordinates $\bar{q}^I$ and $\bar{p}_I$ are not in general canonical with respect to $t$). The way to see this is to realise that the Poisson bracket is the inverse of the symplectic form $\omega$ which is the restriction of the contact form $\omega_{\mathcal{C}}$ to constant time surfaces:
\begin{equation}
\omega=\omega_{\mathcal{C}}|_{t=const},
\end{equation}
which explicitly depends on the choice of the constant time surfaces in $\mathcal{C}$.

\subsection{Group structure}
PTs form a group, denoted by $\mathcal{G}_P$. They include contact transformations, denoted by $\mathcal{G}_C$, as a normal subgroup. Hence, $\mathcal{G}_P$ can be viewed as a principal bundle $\pi: \mathcal{G}_P\mapsto \mathcal{T}$ over the space of all admissible clocks $\mathcal{T}$ with a fibre made of contact transformations, $\mathcal{G}_C$. PTs act simply and transitively in the space of contact coordinates $(q^I,p_I,t)$. Thus, in what follows, $\mathcal{G}_P$ will be identified with the space of contact coordinates. 

\subsection{Special pseudo-canonical transformations}\label{SPT}
Starting from initial contact coordinates $(q^I,~ p_I,~t)$, which is a point in  $\mathcal{G}_P$,  one can define a section\footnote{A section $\sigma$ is an embedding of the base manifold into the bundle such that $\pi\circ\sigma=Id$.} $\sigma: \mathcal{T}\mapsto \mathcal{G}_P$ in the following way: for any choice of time coordinate $\bar{t}$, the respective canonical variables $\bar{q}^I, \bar{p}_I$ are such that Hamilton's equations of motion for the new variables in the new time are obtained by the formal replacement $(q^I,~ p_I,~t)\mapsto (\bar{q}^I,~ \bar{p}_I,~\bar{t})$, that is, the equations of motion:
\begin{equation}\label{sec1}
\frac{\ud q^I}{\ud t}=\frac{\partial \mathrm{h}(q^I,p_I,t)}{\partial p_I},~~\frac{\ud p_I}{\ud t}=-\frac{\partial \mathrm{h}(q^I,p_I,t)}{\partial q^I}
\end{equation}
read in terms of the new variables as: 
\begin{equation}\label{sec2}
\frac{\ud \bar{q}^I}{\ud \bar{t}}=\frac{\partial \mathrm{h}(\bar{q}^I,\bar{p}_I ,\bar{t})}{\partial \bar{p}_I },~~\frac{\ud \bar{p}_I }{\ud \bar{t}}=-\frac{\partial \mathrm{h}(\bar{q}^I,\bar{p}_I ,\bar{t})}{\partial \bar{q}^I}~.
\end{equation}
(Notice a unique formal dependence of the both Hamiltonians on the basic variables and the internal clock). The eqs (\ref{sec1}) and (\ref{sec2}) are {\it physically} equivalent and {\it canonically} inequivalent. If we integrate the equations of motion (\ref{sec1}), we will find constants of motion, denoted by $\mathrm{C}^J(q^I,p_I,t)$. We easily conclude that in the new contact coordinates used in (\ref{sec2}) the constants of motion read $\mathrm{C}^J(\bar{q}^I,\bar{p}_I,\bar{t})$. Hence, $\bar{q}^I, \bar{p}_I$ can be completely fixed by demanding the following set of $2n+1$ algebraic equations to hold:
\begin{equation}\label{cons}
\mathrm{C}^J(q^I,p_I,t)=\mathrm{C}^J(\bar{q}^I,\bar{p}_I,\bar{t}),~~~\bar{t}=\bar{t}(q^I,p_I,t).
\end{equation}
The above relations define a specific family of pseudo-canonical transformations. It can be enlarged if for each contact coordinate system $(\bar{t}, \bar{q}^I, \bar{p}_I)$ derived with (\ref{cons}), one allows for {\it canonical} transformations that move the contact coordinates along a fibre:
\begin{equation}
(\bar{q}^I, \bar{p}_I,\bar{t})\mapsto (\tilde{q}^I, \tilde{p}_I,\bar{t}),
\end{equation}
where $\bar{t}$ is preserved and thus $\pi(\bar{q}^I, \bar{p}_I,\bar{t})=\pi(\tilde{q}^I, \tilde{p}_I,\bar{t})$. Nevertheless, the importance of this particular family of pseudo-canonical transformations given by (\ref{cons}) will become clear when we discuss quantisation. For the moment we just emphasise that Eq. (\ref{cons}) demands that the transformation of contact coordinates is such that the formal dependence of the constants of motion on the contact coordinates is fixed. Nevertheless, any observable that is not a constant of motion will change its form upon the transformation (\ref{cons}).

\subsection{Bianchi type I model}
In what follows we employ the introduced pseudo-canonical transformations (\ref{cons}) to the Bianchi type I model for which the Hamiltonian is given in (\ref{ham00}) and the internal clock is associated with the fluid variable $T$. The reduced phase space is parametrised by $(q,p,k, p_k,\alpha, p_{\alpha})\in\mathbb{R}_+\times\mathbb{R}\times\mathbb{R}_+\times\mathbb{R}\times[0,2\pi)\times\mathbb{R}$.

Transformations to a new clock $\bar{T}$ can be restricted by the use of the so called delay function $D$ of the following form:
\begin{equation}\label{defT}
T\rightarrow\bar{T}=T+D(q,p),
\end{equation}
where we suppress the dependence of $D$ on $(T,k, p_k,\alpha, p_{\alpha})$. Such transformations are general enough for our purposes and at the same time manageable. We find a maximal set of independent constants of motion associated with the Hamiltonian (\ref{ham00}):
\begin{align}\label{motioncons}
\mathrm{C}^1=\frac{c_w^2}{24}p^2-\frac{1}{24}\frac{k^2}{q^2},~~\mathrm{C}^2=qp-\left(\frac{c_w^2}{12}p^2-\frac{1}{12}\frac{k^2}{q^2}\right)T,~~
\mathrm{C}^3=k,\\\nonumber\mathrm{C}^4=p_k-\frac{1}{2c_w}\ln\left(\frac{qp-k/c_w}{qp+k/c_w}\right),~~
\mathrm{C}^5=\alpha,~~\mathrm{C}^6=p_{\alpha},
\end{align}
where the following identifications hold: $\mathrm{C}^1=\mathrm{H}_{true}$ and $\mathrm{C}^2=-2\mathrm{H}_{true}T_0$ ($T_0$ features in eq. (\ref{sol0})). Combining (\ref{defT}) and (\ref{motioncons}) we apply the formula (\ref{cons}) and find new canonical coordinates associated with the clock $\bar{T}=T+D(q,p)$:
\begin{align}\nonumber
\bar{q}=\sqrt{q^2+\frac{c_w^2qpD}{6}+\frac{c_w^2D^2}{6}\left(\frac{c_w^2}{24}p^2-\frac{1}{24}\frac{k^2}{q^2}\right)},~~\bar{p}=\frac{qp+D\cdot\left(\frac{c_w^2}{12}p^2-\frac{1}{12}\frac{k^2}{q^2}\right)}{\sqrt{q^2+\frac{c_w^2qpD}{6}+\frac{c_w^2D^2}{6}\left(\frac{c_w^2}{24}p^2-\frac{1}{24}\frac{k^2}{q^2}\right)}}\\ \label{CT}
\bar{p}_k=p_k+\frac{1}{2c_w}\ln\left(\frac{qp+D\cdot(\frac{c_w^2}{12}p^2-\frac{1}{12}\frac{k^2}{q^2})-k/c_w}{qp+D\cdot(\frac{c_w^2}{12}p^2-\frac{1}{12}\frac{k^2}{q^2})+k/c_w}  \cdot \frac{qp+k/c_w}{qp-k/c_w}\right),~~\bar{k}=k,~~\bar{\alpha}=\alpha,~~\bar{p}_{\alpha}=p_{\alpha}
\end{align}
We note that the constants of motion $k,\alpha,p_{\alpha}$ are preserved by the above transformation as expected. Moreover, for $D\to 0$, the transformation becomes trivial as it should be. By assumption the above transformation preserves the form of the Hamiltonian:
\begin{equation}\label{hamD}
\mathrm{H}_{true,D}=\frac{c_w^2}{24}\bar{p}^2-\frac{1}{24}\frac{\bar{k}^2}{\bar{q}^2},
\end{equation}
which generates exactly the same physical motion in $\mathcal{C}$ as the Hamiltonian (\ref{ham00}), though via a new Poisson bracket with the barred variables forming canonical pairs. 

\begin{figure}[t]
\begin{tabular}{cc}
\includegraphics[width=0.4\textwidth]{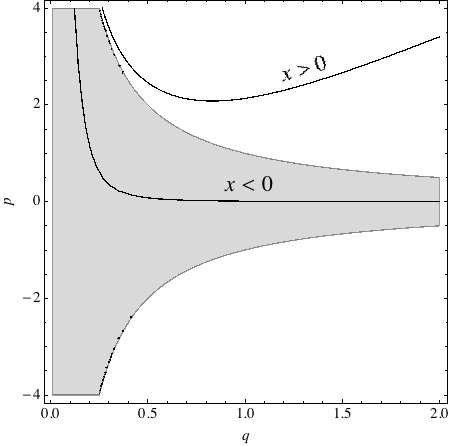}
\hspace{1cm}
\includegraphics[width=0.4\textwidth]{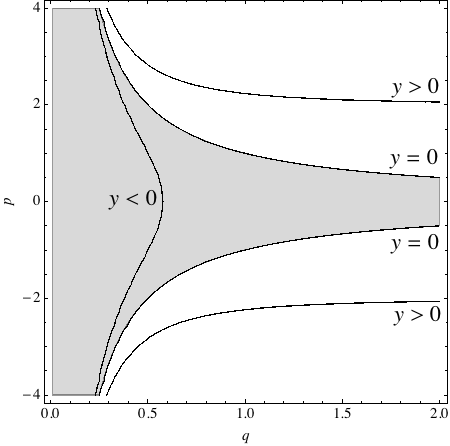}
\end{tabular}
\caption{\small The coordinate system $(x,y)$ in the $(q,p)$-plane.  The level sets of $x$ and $y$ on the left and right, respectively. The shaded region, $y<0$, is classically forbidden.} 
\label{figure3}
\end{figure}

\subsubsection{Delay function}
The old and the new clocks must be monotonic with respect to each other along the motion. Therefore, delay functions must satisfy the following condition:
\begin{equation}\label{conD}
\frac{\ud \bar{T}}{\ud T}=1+\{D(q,p),\mathrm{H}_{true}\}=1+\frac{p}{12}\frac{\partial D}{\partial q} -\frac{k^2}{12q^3}\frac{\partial D}{\partial p}>0
\end{equation}
For facilitating identification of admissible delay functions, we introduce more convenient variables $(x,y)$ in place of $(q,p)$:
\begin{equation}\label{coordtrans}
x=\frac{pq}{p^2-\frac{k^2}{q^2}},~~ y=p^2-\frac{k^2}{q^2},
\end{equation}
where $y\in\mathbb{R}-\{0\}$, $x\in(-\infty,-\frac{k}{y})\cup (\frac{k}{y},\infty)$ for $y>0$ and $x\in(-\frac{k}{|y|},\frac{k}{|y|})$ for $y<0$, see Fig. \ref{figure3}. The transformation of variables is ill-defined for $y=0$. In the new variables, eq. (\ref{conD}) reads
\begin{equation}\label{conD2}
1+\frac{1}{12}\frac{\partial E(x,y)}{\partial x}>0,~~~~E(x(q,p),y(q,p))=D(q,p)
\end{equation}
The variables $x$ and $y$ are useful in defining $D$. Firstly, it is much easier to fulfil the condition (\ref{conD2}) than (\ref{conD}). Secondly, they directly provide the meaning of $D$. Namely, $y$ is a constant of motion and it enumerates both the physical ($y>0$) and non-physical ($y<0$) trajectories. Therefore, the dependence on $y$ tells how a given delay function varies from one trajectory to another one. In particular, it tells whether the clock changes in the physical and non-physical regions of the phase space. Furthermore, for a fixed value of $y$, the variable $x$ varies monotonically along a given trajectory and tells how the clock is modified along the motion. In particular, $x$ tells whether the delay function vanishes at the singularity (see Appendix \ref{C} for further details). Furthermore, the variable $x$ is used to describe the clock transformation far away from the singularity, although, a non-vanishing delay function in the far future/past should not produce the clock effect as the quantum corrections vanish there. We will see this when we come to comparing quantum dynamics in various clocks.

\section{Quantisation and clocks}\label{qc}

\subsection{Quantisation and special pseudo-canonical transformations}\label{points}

In the previous section we discussed a special type of pseudo-canonical transformations to a new clock $t\rightarrow\bar{t}$ and new canonical variables $(q^I,p_I)\rightarrow(\bar{q}^I,\bar{p}_I)$ such that the form of all constants of motion, and in particular of the Hamiltonian, is preserved (see eq. (\ref{cons})). It follows that it is sufficient to solve the Hamilton equations in one clock (\ref{sec1}) to obtain the solution in all the other clocks (\ref{sec2}). Namely, given the solution in $t$ in the form
 \begin{align}q^I=Q^I(t,C^1,\dots,C^{2n}),~~p_I=P_I(t,C^1,\dots,C^{2n}),\end{align}
where $C^1,\dots,C^{2n}$ are constants of motion which parametrise the space of solutions, we obtain the solution in $\bar{t}$ by the simple replacement of clock and canonical variables: 
\begin{align}\bar{q}^I=Q^I(\bar{t},C^1\dots,C^{2n}),~~\bar{p}_I=P_I(\bar{t},C^1,\dots,C^{2n}).\end{align} 
By assumption, the constants $C^1,\dots,C^{2n}$ are independent of the choice of clock.

The formal correspondence between canonical formalisms related by special pseudo-canonical transformations can be extended to the quantum level. Let us consider `quantisation' by which we mean a linear map, $$\mathcal{Q}: C^{\infty}(\mathbb{P})\mapsto \mathcal{L}(\mathcal{H}),$$ from phase space functions to linear operators in the Hilbert space $\mathcal{H}$. Let $q^I,~p_I$ be canonical coordinates and $\mathrm{O}(q^I,~p_I)$ be any observable, and let
\begin{align}
\mathcal{Q}:~ \mathrm{O}(q^I,~p_I)\mapsto\hat{\mathrm{O}}.
\end{align}
The same quantisation map $\mathcal{Q}$ may be applied to another reduced phase space based on another clock $\bar{t}$ and parametrised by canonical coordinates $\bar{q}^I,~\bar{p}_I$. Namely, let an observable $\mathrm{O}(\bar{q}^I,~\bar{p}_I)$ which depends formally in the same way on the respective basic variables as $\mathrm{O}(q^I,~p_I)$ be promoted to the same operator,
\begin{align}\label{eq}
\mathcal{Q}:~ \mathrm{O}(\bar{q}^I,~\bar{p}_I)\mapsto\hat{\mathrm{O}}.
\end{align}
In this way, a single quantisation map $\mathcal{Q}$ is sufficient to quantise all canonical formalisms related by special pseudo-canonical transformations given by eq. (\ref{cons}) and it establishes all the respective quantum theories in the same Hilbert space $\mathcal{H}$. Let us discuss the properties of the obtained quantum theories:

{\bf 1.} First of all we notice that observables $\mathrm{O}(q^I,~p_I)$ and $\mathrm{O}(\bar{q}^I,~\bar{p}_I)$, though promoted to the same linear operator by the map $\mathcal{Q}$, they in general represent physically different quantities. Thus, provided that $\mathcal{Q}$ is an injection, it immediately follows that a unique physical quantity is in general assigned a different operator if quantised in a different canonical formalism.

{\bf 2.} The exceptional observables are constants of motion, $C^J(q^I,~p_I,t)$ and $C^J(\bar{q}^I,\bar{p}_I,\bar{t})$, which are mapped by $\mathcal{Q}$ to the same quantum operator in $\mathcal{H}$ and which share the same physical meaning by the virtue of Eq. (\ref{cons}). Therefore, the operators $\hat{C}^J$ (or, the families of operators $\hat{C}^J(\tau)$, where $\tau=t$ or $\tau=\bar{t}$) represent the same conserved physical quantities for all quantum theories (in all internal clocks).

{\bf 3.} Since the Hamiltonian is a constant of motion, it is assigned by $\mathcal{Q}$ a unique quantum operator for all internal clocks. Hence, for any initial vector state $|\psi(\tau_0)\rangle\in\mathcal{H}$, it generates via the Schr\"odinger equation,
\begin{align}i\frac{\partial}{\partial \tau}|\psi\rangle=\hat{H}|\psi\rangle,\end{align}
a curve in the Hilbert space, $\tau\mapsto |\psi(\tau)\rangle\in\mathcal{H}$, which is unique for all clocks up to parametrisation, $\tau=t$ or $\tau=\bar{t}$.

{\bf 4.} Suppose $\{\hat{C}^j\},~j\in J$ is a maximal set of quantised constants of motions which include the Hamiltonian, are independent, commute with each other and are self-adjoint. Then by the virtue of spectral theorem, any vector state $|\psi\rangle\in\mathcal{H}$ can be expressed as a sum of the eigenstates of ${\hat{C}^j}$'s. In other words, it can be represented as a wavefunction on the spectra of ${\hat{C}^j}$'s, 
\begin{align}\psi(\dots,c^j,\dots)=\langle \dots,\phi_{c^j},\dots |\psi\rangle,\end{align}
where 
\begin{align}\hat{C}^j|\phi_{c^j}\rangle=c^j\cdot |\phi_{c^j}\rangle.\end{align}
Since the constants of motion have the same physical meaning in all internal clocks, we conclude that $\psi(\dots,c^j,\dots)$ and the associated probability distribution has a unique physical interpretation for all clocks.

{\bf 5.} By an analogous reasoning, we conclude that the spectral decomposition of any vector state $|\psi\rangle\in\mathcal{H}$ in terms of a wavefunction on the spectra of a maximal set of commuting quantised observables which are not constants of motion
\begin{align}\psi(\dots,d^j,\dots)=\langle \dots,\varphi_{d^j},\dots |\psi\rangle,\end{align}
where $d^j$ stands for an eigenvalue of a dynamical operator $\hat{D}^j$,
\begin{align}\hat{D}^j|\varphi_{d^j}\rangle=d^j\cdot |\varphi_{d^j}\rangle,\end{align} 
cannot be given a unique physical interpretation for all internal clocks. It is so simply because $d^j$'s have different physical meanings in different internal clocks. In other words, dissimilarities between {\it dynamical} properties of vector states seen in different internal clocks are unavoidable. For instance, if $|\psi\rangle$ is a state of a moving particle, then its position is a dynamical observable and the probability distribution of the particle's position in the state $|\psi\rangle$ depends on the chosen internal clock. It is clear that the dissimilarities have nothing to do with usual quantisation ambiguities (like ambiguous orderings, etc) since a unique quantisation map $\mathcal{Q}$ is employed for all internal clocks.

\subsection{Bianchi Type I model}
A quantum theory of the Bianchi Type I model was developed in \cite{BI}. The essential elements of the quantisation procedure are given below. The distinctive feature of the studied model is the phase space which consists of the physical ($\mathrm{H}_{true}>0$) and non-physical ($\mathrm{H}_{true}<0$) regions. For isotropic variables $(q,p)$ we employ an integral quantisation based on coherent states which are obtained with a unitary irreducible representation of the affine group. For the remaining variables we use canonical quantisation, although the respective quantum operators will not play a crucial role in our analysis.

\subsubsection{Coherent states}
Let us first introduce the coherent states which will play a key role in establishing the quantum theory. They will be applied to the isotropic variables $(q,p)\in\mathbb{R}_+\times\mathbb{R}$. The affine group is a minimal group of canonical transformations defined as 
\begin{align}(\mathbb{R}_+\times\mathbb{R})\circ (\mathbb{R}_+\times\mathbb{R})\ni (q',p')\circ (q,p)\mapsto \left(qq',\frac{p}{q'}+p'\right)\in\mathbb{R}_+\times\mathbb{R},\end{align}
in the phase space $(q,p)$ equipped with the symplectic form $\omega=\ud q\ud p$. The following family of unitary operators
\begin{align}U(q,p)\cdot\psi(x)=e^{ipx}\psi(x/q)\in L^2(\mathbb{R}_+,\ud x)\end{align}
are easily verified to form an irreducible representation of the affine group,
\begin{align}U(q',p')\cdot U(q,p)=U(qq',\frac{p}{q'}+p').
\end{align} 
The affine coherent states are defined as a continuous map from the phase space to the Hilbert space,
\begin{align}\label{ACS}
\mathbb{R}_+\times\mathbb{R}\ni(q,p)\mapsto |q,p\rangle:=e^{ipx}\psi(x/q)\in\mathcal{H}=L^2(\mathbb{R}_+,\ud x),
\end{align}
where $\psi(x)\in L^2(\mathbb{R}_+,\ud x/x)\cap L^2(\mathbb{R}_+,\ud x)$, called the fiducial vector, is fixed and some possible choices for $\psi(x)$ are given in eq. (\ref{fiducial}) of Appendix \ref{D}. The coherent states form an overcomplete continuous basis in $\mathcal{H}$,
\begin{align}\label{unity}
\int_{\mathbb{R}_+\times\mathbb{R}}\frac{\ud q\ud p}{2\pi c_{-1}} |q,p\rangle\langle q,p|=\mathbf{1},
\end{align}
where $c_{-1}=\int_0^{\infty} \frac{|\psi(x)|^2}{x}~\ud x$. The resolution of unity (\ref{unity}) is the essential property of coherent states and is guaranteed by the construction based on a unitary irreducible group \cite{JPt}.

\subsubsection{Quantisation map} 

The quantisation map for phase space functions of $q$, $p$ is based on the affine coherent states and reads:
\begin{align}\label{Qqp}
\mathcal{Q}:~O(q,p)\mapsto \hat{O}=\int_{\mathbb{R}_+\times\mathbb{R}}\frac{\ud q\ud p}{2\pi c_{-1}} O(q,p)|q,p\rangle\langle q,p|
\end{align}
The above map is linear, it assigns the identity to $O(q,p)=1$ and it assigns positive operators to positive functions. This quantisation is very convenient in imposing the positivity constraint which is satisfied by the classical model (see below). Moreover, it respects the affine symmetry of the phase space and leads to the singularity resolution as shown shortly. Nevertheless, the use of this specific quantisation map is completely irrelevant for the existence of ambiguity associated with different physical interpretations as described in the subsection (\ref{points}) due to the choice of internal clock.

The quantisation of phase space functions of the remaining variables $k,p_k,\alpha,p_{\alpha}$ is {\it canonical}, i.e.
\begin{align}
\mathcal{Q}:~~k\mapsto  \hat{k}=k,~~~p_k\mapsto \hat{p}_k=-i\partial_{k},~~ \alpha\mapsto\hat{\alpha}=\alpha,~~~~p_{\alpha}\mapsto\hat{p}_{\alpha}=-i\partial_{\alpha},
\end{align}
and compound functions of the above elementary variables are promoted to respective compound operators which are symmetric with respect to the above elementary operators.

\subsubsection{Quantum Hamiltonian} We will use the map $\mathcal{Q}$ defined above to quantise the Hamiltonian (\ref{ham00}), that is,
$$\mathrm{H}_{true}=\frac{c_w^2}{24}p^2-\frac{1}{24}\frac{k^2}{q^2},$$ 
together with the positivity constraint, $\mathrm{H}_{true}>0$. The variable $k$ is simply re-interpreted as a multiplication operator on $L^2(\mathbb{R}_+,\ud k)$ upon canonical quantisation. We are going to treat it as a constant and treat $\mathrm{H}_{true}$ as a function of $q,p$. We impose the positivity constraint on the quantum theory by quantising 
\begin{align}\theta(\mathrm{H}_{true})\mathrm{H}_{true}\end{align} 
instead of $\mathrm{H}_{true}$ alone, where $\theta(\cdot)$ is the Heaviside function. Note that the classical Hamiltonians $\theta(\mathrm{H}_{true})\mathrm{H}_{true}$ and $\mathrm{H}_{true}$ are equal in the physical region $\mathrm{H}_{true}>0$ and different in the non-physical region $\mathrm{H}_{true}<0$. The choice of classical Hamiltonian (as long as it is valid) as a starting point for quantisation of dynamics is free.  This choice is usually guided by the kind of quantum theory that one expects to get. We notice that $\theta(\mathrm{H}_{true})\mathrm{H}_{true}$ becomes trivial in the unphysical region and does not generate spurious dynamics there. Incorporating the quantum uncertainties may result in weakening the classical constraint and induce some dynamics in the classically forbidden region. The quantum Hamiltonian reads:
\begin{align}\label{QHBI}
\mathcal{Q}:~\theta(\mathrm{H}_{true})\mathrm{H}_{true}\mapsto \hat{\mathrm{H}}_{true}=\int_{\mathbb{R}_+\times\mathbb{R}}\frac{\ud q\ud p}{2\pi c_{-1}} \theta(\mathrm{H}_{true})\mathrm{H}_{true}|q,p\rangle\langle q,p|.
\end{align}
See Appendix \ref{D} for the result of evaluation of the above integral. The final formula is very difficult to analyse and therefore, we will employ phase space portraits to represent the quantum dynamics only approximately.
\subsubsection{Switching to other internal clocks}
The form of the Hamiltonian $\mathrm{H}_{true}(q,p)$ is preserved upon special pseudo-canonical transformations to new contact coordinates $(t,q,p)\mapsto (\bar{t},\bar{q},\bar{p})$ discussed in the subsection (\ref{SPT}). Thus, the form of $\theta(\mathrm{H}_{true})\mathrm{H}_{true}$ is also preserved and the quantum Hamiltonian of eq. (\ref{QHBI}) is common for all internal clocks on the basis of the discussion around eq. (\ref{eq}). Therefore, for any initial state $|\psi\rangle$ its evolution is the same in all internal clocks (see point 3. in the subsection (\ref{points})). 

The interpretation of any state $|\psi\rangle$ in general depends on the employed clock. Let us illustrate it with an example. The variables $q$ and $\bar{q}$ are assigned the same operator by $\mathcal{Q}$, namely
\begin{align}\mathcal{Q}:~q\mapsto \hat{q}=\int_{\mathbb{R}_+\times\mathbb{R}}\frac{\ud q\ud p}{2\pi c_{-1}}q~ |q,p\rangle\langle q,p|=\int_{\mathbb{R}_+\times\mathbb{R}}\frac{\ud \bar{q}\ud \bar{p}}{2\pi c_{-1}}\bar{q}~ |\bar{q},\bar{p}\rangle\langle \bar{q},\bar{p}|=\hat{\bar{q}}\end{align}
(where we have applied the formula (\ref{Qqp})). However, they represent different dynamical quantities as shown by eq. (\ref{CT}). If we solve the eigenvalue problem, $\hat{q}\phi_{\lambda}=\lambda\phi_{\lambda}$ (or $\hat{\bar{q}}\phi_{\lambda}=\lambda\phi_{\lambda}$), then we can represent any state $|\psi\rangle\in\mathcal{H}$ as a wavefunction on the spectrum of $\hat{q}$ (and $\hat{\bar{q}}$),
\begin{align}\psi(\lambda)=\langle\phi_{\lambda}|\psi\rangle,\end{align}
which is a unique wavefunction. However, the interpretation of the wavefunction is different in different internal clocks as the interpretation of its argument $\lambda$ corresponds to $q$ or to $\bar{q}$ depending on the internal clock.  

Alternatively, we may notice that the operator
\begin{align}\mathcal{Q}:~\bar{q}\mapsto \hat{\bar{q}}=\int_{\mathbb{R}_+\times\mathbb{R}}\frac{\ud \bar{q}\ud \bar{p}}{2\pi c_{-1}}\bar{q}~ |\bar{q},\bar{p}\rangle\langle \bar{q},\bar{p}|\end{align}
in the internal clock $\bar{t}$ corresponds to the same physical quantity as the operator
\begin{align}\mathcal{Q}:~\bar{q}(q,p)\mapsto \widehat{\bar{q}(q,p)}=\int_{\mathbb{R}_+\times\mathbb{R}}\frac{\ud q\ud p}{2\pi c_{-1}}\bar{q}(q,p)~ |q,p\rangle\langle q,p|\end{align}
in the internal clock $t$, where $\bar{q}=\bar{q}(q,p)$ is a compound function of $q$ and $p$ derived in eq. (\ref{CT}). But now the operators $\hat{\bar{q}}\phi_{\lambda}=\lambda\phi_{\lambda}$ and $\widehat{\bar{q}(q,p)}\varphi_{\lambda}=\lambda\varphi_{\lambda}$ have different eigenvalue solutions despite the fact that they correspond to the same physical quantity. So any state vector $|\psi\rangle\in\mathcal{H}$ will be given a different wavefunction of the argument $\lambda$ corresponding the same physical quantity $\bar{q}$,
\begin{align}\langle\phi_{\lambda}|\psi\rangle=f(\lambda)\neq g(\lambda)=\langle\varphi_{\lambda}|\psi\rangle ,\end{align}
depending on the employed clock. Thus, the physical interpretation of the state $|\psi\rangle$ is clearly different for different clocks. Performing this kind of comparison explicitly is cumbersome as, for instance, the operator corresponding to $\bar{q}(q,p)$ may be too complex to determine its eigenstates. In the next section we propose a more efficient way to tackle the problem of comparing descriptions of quantum dynamics in different clocks.

\section{Phase space portraits}
\label{spsp}
The goal of this section is to present a semiclassical approximation to quantum dynamics based on phase space portraits and then to apply it to the quantised Bianchi-I model. This approximation is useful because it is significantly simpler to obtain than the full quantum description. On one hand, it includes only the crudest corrections from quantum theory to the classical dynamics. On the other hand, if any dissimilarities between different clock-based quantum dynamics are found, then going to a more detailed description will not make the dissimilarities go away. On the contrary, the extent of dissimilarities can only enlarge.

\subsection{Klauder's approach}

Let us recall Klauder's approach \cite{klauderscm,BI} to semiclassical approximation to quantum motion. Let the `quantum action' be defined as:
\begin{equation}\label{action}
S(\psi,\dot{\psi})=\int\lg\psi|i\frac{\partial}{\partial t} - \hat{\mathrm{H}}|\psi\rg\ud t,
\end{equation}
where $|\psi\rangle\in\mathcal{H}$ belongs to the Hilbert space and $\hat{\mathrm{H}}\in\mathcal{L}(\mathcal{H})$ is the quantum Hamiltonian. Minimisation of the quantum action with respect to variations of $\psi$ (or $\psi^{*}$) leads to the Schr\"odinger equation,
$$i\frac{\partial}{\partial t}|\psi\rg= \hat{\mathrm{H}}|\psi\rg,$$
which defines the exact quantum motion. The semiclassical approximation can be derived by confining the quantum motion to a family coherent states defined in (\ref{ACS}). By setting
 $$|\psi\rg\in\{|q,p\rg:(q,p)\in\mathbb{R}_+\times\mathbb{R}\},$$
the quantum action (\ref{action}) reads
\begin{equation}\label{action2}
S(q,\dot{q},p,\dot{p})=\int\lg q,p|i\frac{\partial}{\partial t} - \hat{\mathrm{H}}|q,p \rg\ud t=\int\left(\dot{q}p-\lg q,p|\hat{\mathrm{H}}|q,p \rg\right)\ud t
\end{equation}
The minimisation of (\ref{action2}) with respect to variations of $q$ and $p$ leads to the Hamilton equations for semiclassical motion,
\begin{equation}\label{SME}
\frac{\ud q}{\ud t}=\frac{~~\partial \check{\mathrm{H}}}{\partial p}~,~~\frac{\ud p}{\ud t}=-\frac{~~\partial \check{\mathrm{H}}}{\partial q}~,~~\check{\mathrm{H}}(q,p)=\langle q,p|\hat{\mathrm{H}}|q,p\rangle .
\end{equation}
The solution to the above equations gives the phase space portrait of quantum dynamics. It is simple to notice that starting from another reduced phase space with clock $\bar{t}$, quantisation and semiclassical approximation must lead to formally the same Hamilton equations of semiclassical motion, namely
\begin{equation}
\frac{\ud\bar{q}}{\ud\bar{t}}=\frac{~~\partial \check{\mathrm{H}}}{\partial \bar{p}}~,~~\frac{\ud\bar{p}}{\ud\bar{t}}=-\frac{~~\partial \check{\mathrm{H}}}{\partial \bar{q}}~,~~\check{\mathrm{H}}(\bar{q},\bar{p})=\langle \bar{q},\bar{p}|\hat{\mathrm{H}}|\bar{q},\bar{p}\rangle .
\end{equation}
This is so because (i) the classical Hamiltonian is a unique function of basic variables irrespectively of the employed clock; (ii) the quantisation map (\ref{Qqp}) promotes it to a unique quantum operator irrespectively of the employed clock; (iii) we use a unique family of coherent states with renamed labels $(q,p)\mapsto(\bar{q},\bar{p})$ to approximate the quantum motion.
\subsection{Comparison of phase space portraits}
It was demonstrated in the previous section that the quantum motion is unique for all internal clocks. The quantisation map $\mathcal{Q}$ of (\ref{Qqp}) assigns the same operators to the canonical pairs $q$, $p$ and $\bar{q}$, $\bar{p}$ because in the definition of  $\mathcal{Q}$ a unique family of coherent states is implicitly assumed, i.e.
\begin{align}\forall_{q=\bar{q}}~\forall_{p=\bar{p}}~~|q,p\rangle=|\bar{q},\bar{p}\rangle.\end{align}
Hence, confining the quantum motion to $|q,p\rangle$'s is identical with confining it to $|\bar{q},\bar{p}\rangle$'s because they are in fact the same coherent states. Since the exact dynamics is unique, the semiclassical dynamics understood as an approximate quantum motion, 
\begin{align}\mathbb{R}\ni t\mapsto |q(t),p(t)\rg\in\mathcal{H},\end{align}
must be unique as well. So, where are the dissimilarities?

As already discussed, any state $|\psi\rg\in\mathcal{H}$ has a physical interpretation in terms of a wavefunction on the eigenvalues of any quantum (self-adjoint) operator that corresponds to a physical observable. A limited physical interpretation of the state $|\psi\rg$ can be given e.g. by the expectation value of a given operator in the state $|\psi\rg$ rather than a complete wavefunction. If a given quantum operator is dynamical then the physical observable to which it corresponds  depends on the internal clock. For example, the coherent states,
$$\mathcal{H}\ni |a,b\rangle, ~(a,b)\in\mathbb{R_+}\times\mathbb{R},$$
can be given a physical interpretation in terms of the expectation values of the operators $\hat{q}$ and  $\hat{p}$ or $\hat{\bar{q}}$ and $\hat{\bar{p}}$ depending on the chosen internal clock. It can be verified that
\begin{align}\langle a,b|\hat{q}|a,b\rangle=\langle a,b|\hat{\bar{q}}|a,b\rangle=a,\\\nonumber
\langle a,b|\hat{p}|a,b\rangle=\langle a,b|\hat{\bar{p}}|a,b\rangle=b\end{align}
In other words, for the clock $t$, the above coherent state is interpreted as a semiclassical state of the system corresponding to
\begin{align}q=a,~p=b,\end{align}
where $q$ and $p$ are basic dynamical observables. Whereas for the clock $\bar{t}$ it describes a semiclassical state corresponding to
\begin{align}\bar{q}=a,~\bar{p}=b,\end{align}
where $\bar{q}$ and $\bar{p}$ are another set of basic dynamical observables. Taking into account a non-trivial relation $\bar{q}=\bar{q}(q,p)$ and $\bar{p}=\bar{p}(q,p)$ given eq. (\ref{CT}), we have just proved a dissimilarity in the interpretation of the state $|a,b\rangle$ in terms of basic dynamical variables.

The method of comparison which we employ below is based on the phase space portraits. We first derive the phase space portraits of quantum dynamics with different clocks and in different reduced phase spaces. Then we employ the coordinate relation between the reduced phase spaces (\ref{cons}), 
$$
\mathrm{C}^J(q^I,p_I,t)=\mathrm{C}^J(\bar{q}^I,\bar{p}_I,\bar{t}),~~~\bar{t}=\bar{t}(q^I,p_I,t),
$$
(for the Bianchi-I the relation is given in eq. (\ref{CT})) to pullback the phase space portrait in $(\bar{q}^I,\bar{p}_I)$ to the phase space portrait in $(q^I,p_I)$ via $(\bar{q}^I,\bar{p}_I)\mapsto (q^I(\bar{q}^I,\bar{p}_I),p_I(\bar{q}^I,\bar{p}_I))$. Finally, the two semiclassical motions are compared in a fixed reduced phase space $(q^I,p_I)$ and dissimilarities are discussed.

\subsection{Phase space portrait of Bianchi type I model} 
The Klauder-like approximation to the quantum Bianchi-I dynamics given by the Hamiltonian (\ref{QHBI}) was derived in \cite{BI}. In what follows we recall the most essential result. The semiclassical Hamiltonian is defined
\begin{equation}\check{\mathrm{H}}_{true}=\langle q,p| \hat{\mathrm{H}}_{true}|q,p\rangle=\int_{\mathbb{R}_+\times\mathbb{R}}\frac{\ud q'\ud p'}{2\pi c_{-1}} \theta(\mathrm{H}_{true})\mathrm{H}_{true}|\langle q',p'|q,p\rangle|^2 .\end{equation}
(which is the expectation value of $\hat{\mathrm{H}}_{true}$ in the coherent states $|q,p\rangle$). The explicit determination of the above integral gives:
\begin{align}\label{semiH}
\check{\mathrm{H}}_{true}=\frac{c_w^2}{24}\left(p^2-\frac{B(\frac{k}{c_w})^2}{q^2}\frac{\lambda_2(qp)^2}{1+\lambda_1(\frac{k}{c_w})^2+\lambda_2(qp)^2}+\frac{A}{q^2}\frac{1+\lambda_2(qp)^2}{1+\lambda_1(\frac{k}{c_w})^2+\lambda_2(qp)^2}\right),
\end{align}
where the values of parameters $A,~B,~\lambda_1$ and $\lambda_2$ depend on the specific family of coherent states via the fiducial vector $\psi(x)$ used to define the coherent states $|q,p\rangle$ in (\ref{ACS}). The explicit values of these parameters are irrelevant for our purposes and can be found in \cite{BI}\footnote{The parameters used in \cite{BI} are rescaled here for brevity as follows: $\frac{B}{\lambda^2}\rightarrow B$, $\frac{A}{\lambda^2}\rightarrow A$ and $\lambda_2\lambda^2\rightarrow \lambda_2$. In effect, $\lambda$ is absent in our present notation.}. In order to make the plots of the semiclassical dynamics, we employ the fiducial vector of Eq. (\ref{fiducial}) with $\nu=3$ to fix the above parameters.

The semiclassical dynamics generated by (\ref{semiH}) is explicitly integrable, though involved. We find:
\begin{align} \label{semisol1}
q=\sqrt{\frac{1}{\check{\mathrm{H}}_{true}}\frac{c_w^2}{24}\left((2\check{\mathrm{H}}_{true}T)^2-\frac{B(\frac{k}{c_w})^2\lambda_2(2\check{\mathrm{H}}_{true}T)^2}{1+\lambda_1(\frac{k}{c_w})^2+\lambda_2(2\check{\mathrm{H}}_{true}T)^2}+A\frac{1+\lambda_2 (2\check{\mathrm{H}}_{true}T)^2}{1+\lambda_1 (\frac{k}{c_w})^2+\lambda_2(2\check{\mathrm{H}}_{true}T)^2}\right)}\\ \label{semisol12}
p=\frac{2\check{\mathrm{H}}_{true} T}{q},~~
k=const,~~\alpha=const,~~p_{\alpha}=const,\\ \label{semisol2}
p_k=A_0\arctan(\check{\mathrm{H}}_{true}\omega_0T)+\frac{A_1}{\omega_+}\arctan(\check{\mathrm{H}}_{true}\omega_+T)-\frac{A_1}{\omega_-}\arctan(\check{\mathrm{H}}_{true}\omega_-T)+const,
\end{align}
for the bounce assumed to take place at $T=0$. The values of constant parameters $\omega_0$, $\omega_{\pm}$ and $A_0$ are given in Appendix \ref{B}.

\begin{figure}[t]
\begin{tabular}{cc}
\includegraphics[width=0.4\textwidth]{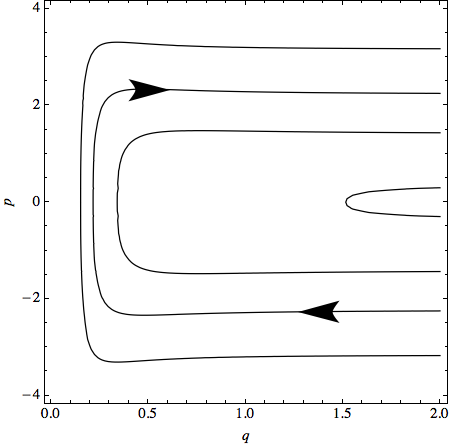}
\hspace{1cm}
\includegraphics[width=0.4\textwidth]{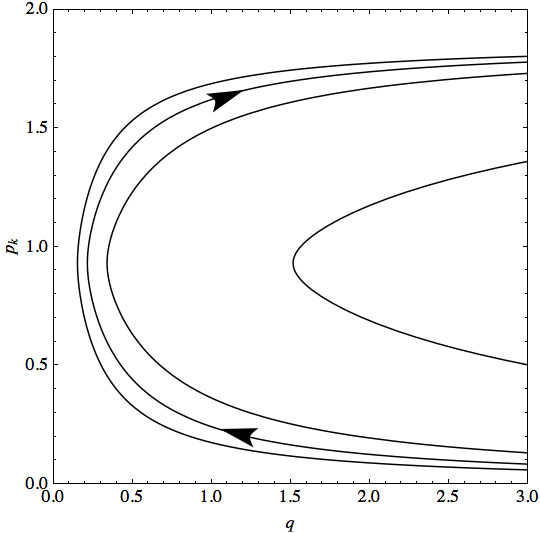}
\end{tabular}
\caption{\small Phase space portrait of the semiclassical dynamics in the planes $(q,p)$ and $(q,p_k)$ for $k=1$, $c_w=1$ and $\check{\mathrm{H}}_{true}=\frac{1}{240},\frac{1}{12},\frac{5}{24}, \frac{5}{12}$. All the phase space is accessible by the semiclassical dynamics. The trajectories do not diverge but reverse. They represent bouncing universes.} 
\label{figure2}
\end{figure}

It is clear from (\ref{semisol1}) that the singularity is removed and the minimal value of $q$ reads 
\begin{align}q_{min}=\sqrt{\frac{c_w^2A}{24\check{\mathrm{H}}_{true}\left(1+\lambda_1 (\frac{k}{c_w})^2\right)}}.\end{align}
The solutions exhibit a bounce: a contracting semiclassical universe crosses the classically forbidden phase space region, $\mathrm{H}_{true}<0$, and smoothly transforms into an expanding universe, as depicted in Fig \ref{figure2}. Crossing the classically forbidden region is necessary to connect the expanding and contracting classical solutions that are topologically disconnected by this region in the kinematical phase space.

We infer from our previous discussions that in order to quantise dynamics and next approximate it for other reduced phase spaces one needs to apply the unique quantisation formula (\ref{QHBI}) and the unique formula for the semiclassical Hamiltonian (\ref{SME}) by substituting the variables $(q,p)$ with $(\bar{q},\bar{p})$. As a result, the semiclassical Hamiltonian in other reduced phase spaces reads
\begin{align}\label{hamDsemi}
\check{\mathrm{H}}_{D,true}=\frac{c_w^2}{24}\left(\bar{p}^2-\frac{B(\frac{\bar{k}}{c_w})^2}{\bar{q}^2}\frac{\lambda_2(\bar{q}\bar{p})^2}{1+\lambda_1(\frac{\bar{k}}{c_w})^2+\lambda_2(\bar{q}\bar{p})^2}+\frac{A}{\bar{q}^2}\frac{1+\lambda_2(\bar{q}\bar{p})^2}{1+\lambda_1(\frac{\bar{k}}{c_w})^2+\lambda_2(\bar{q}\bar{p})^2}\right)
\end{align}
that is, it is formally equivalent with the Hamiltonian of Eq. (\ref{semiH}).

\subsection{Comparing phase space portraits for quantum Bianchi type I model}

We employ different internal clocks to derive semiclassical phase space portraits for the unique quantum dynamics of Bianchi type I model. Next we compare them in a single reduced phase space. We skip the variables $\alpha$, $p_{\alpha}$ and $k$ since they are constants of motion and no clock effect can occur for them. We focus on the space $(q,p,p_k)$.

More precisely, we fix a single reduced phase space parametrised by the barred variables $(\bar{q},\bar{p},\bar{p}_k)$ and the remaining reduced phase spaces parametrised by $(q,p,p_k)$. In order to arrive at the fixed reduced phase space $(\bar{q},\bar{p},\bar{p}_k)$, a pseudo-canonical transformation (\ref{CT}) based on an appropriate delay function $D(q,p)$ is performed on $(q,p,p_k)$.

The scheme of computations is following. We use the solution (\ref{semisol1},\ref{semisol12},\ref{semisol2}) to the semiclassical Hamilton equations generated by the Hamiltonian (\ref{semiH}):
\begin{equation}\label{solqppk}T\mapsto (q(T),p(T),p_k(T)).\end{equation}
Next we perform a pseudo-canonical transformation with a suitable delay function $D(q,p)$, described by the eqs (\ref{CT}). Combining it with (\ref{solqppk}), we obtain: 
\begin{align}T\mapsto (\bar{q}(T),\bar{p}(T),\bar{p}_k(T)).\end{align}
The reduced phase space $(\bar{q},\bar{p},\bar{p}_k)$ is where we make the comparison of semiclassical dynamics descendent from reduced phase space based on internal clocks that differ by delay functions $D(q,p)$. Of course, the semiclassical dynamics which is derived directly in $(\bar{q},\bar{p},\bar{p}_k)$:
 \begin{align}\bar{T}\mapsto (\bar{q}(\bar{T}),\bar{p}(\bar{T}),\bar{p}_k(\bar{T})),\end{align}
is already represented in Fig. \ref{figure2} and will serve as a basis for the comparisons.


\section{Numerical examples}\label{numex}

In what follows we study four examples of delay functions and the effect that the respective internal clocks take on the semiclassical phase space portraits. The admissible delay functions were given in eq. (\ref{conD2}). Because this particular model suffers from a singularity at the classical level, the extra demand that we make is that the delay functions vanish at the singularity (for technical details see Appendix \ref{C}). 

For convenience we introduce the following function:
\begin{align}\eta(z)=\theta(-z)e^{\frac{1}{z}}\in[0,1),~~z\in\mathbb{R}\end{align}

\begin{figure}[t]
\centering
\includegraphics[width=.38\textwidth]{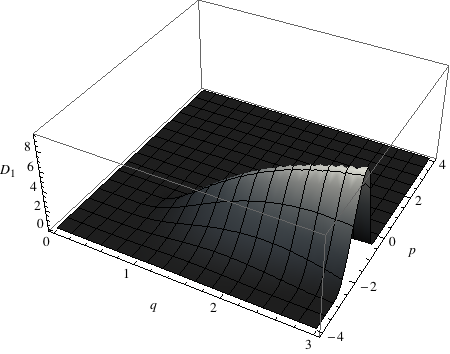}
\includegraphics[width=.3\textwidth]{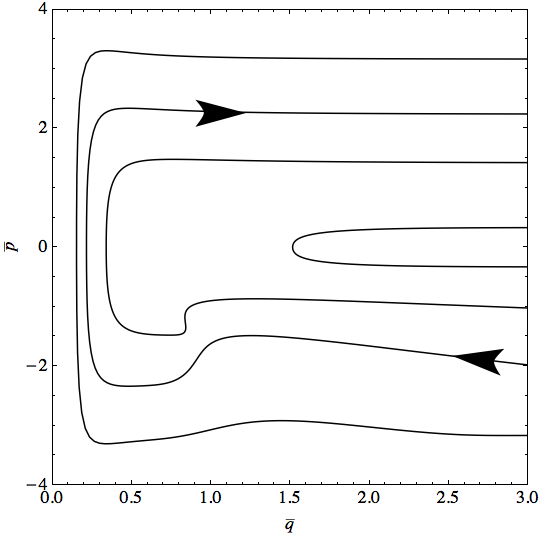}
\includegraphics[width=.3\textwidth]{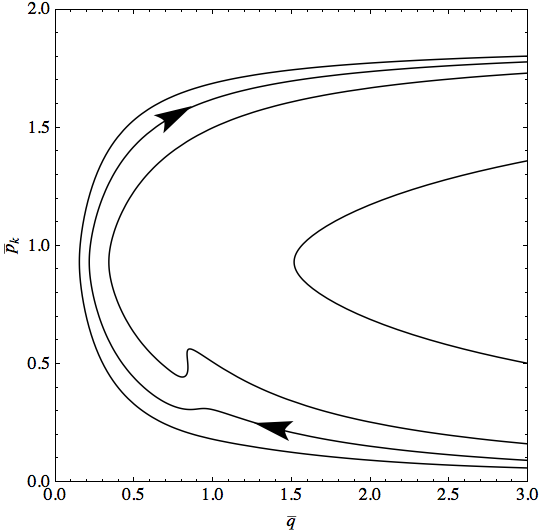}
\caption{\small From left to right: (i) plot of the delay function $D_1$ which defines the transformation between the reduced phase space used to derive the semiclassical dynamics and the reduced phase space used to describe it; (ii) the semiclassical evolution in the $(\bar{q},\bar{p})$-plane for $k=1$, $c_w=1$, $\check{\mathrm{H}}_{true}=\frac{1}{240},\frac{1}{12},\frac{5}{24}, \frac{5}{12}$; (iii) the semiclassical evolution in the $(\bar{q},\bar{p}_k)$-plane.}
\label{d1}
\end{figure}

\begin{figure}[t]
\centering
\includegraphics[width=.38\textwidth]{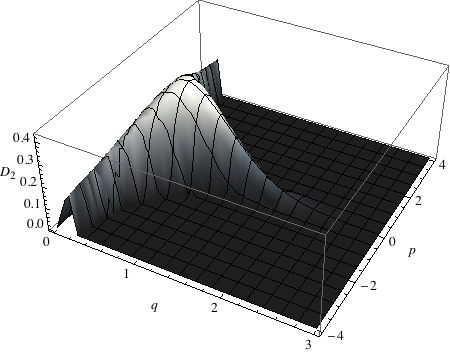}
\includegraphics[width=.3\textwidth]{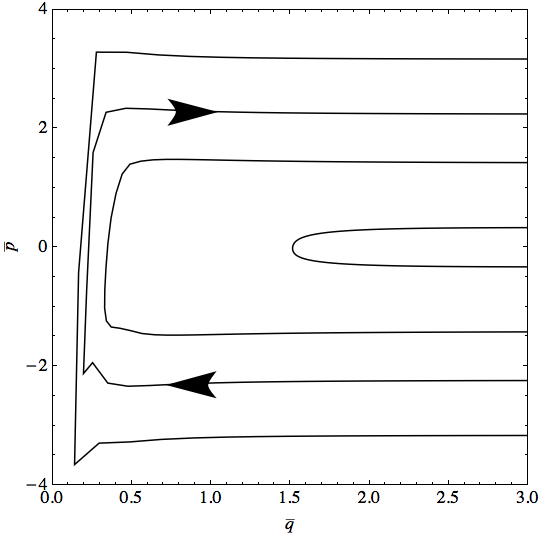}
\includegraphics[width=.3\textwidth]{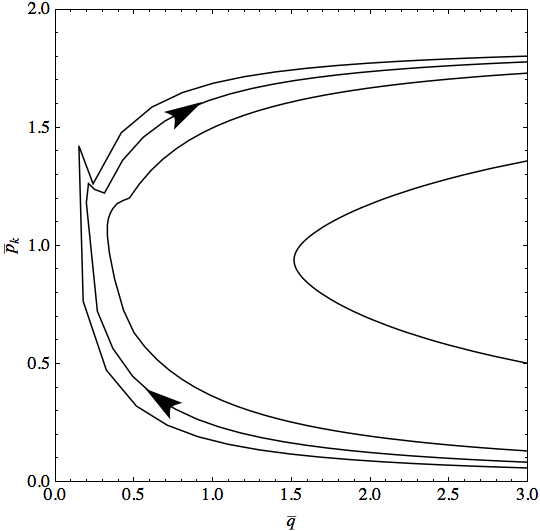}
\caption{\small From left to right: (i) plot of the delay function $D_2$ which defines the transformation between the reduced phase space used to derive the semiclassical dynamics and the reduced phase space used to describe it; (ii) the semiclassical evolution in the $(\bar{q},\bar{p})$-plane for $k=1$, $c_w=1$, $\check{\mathrm{H}}_{true}=\frac{1}{240},\frac{1}{12},\frac{5}{24}, \frac{5}{12}$; (iii) the semiclassical evolution in the $(\bar{q},\bar{p}_k)$-plane.} 
\label{d2}
\end{figure}

\begin{figure}[t]
\centering
\includegraphics[width=.38\textwidth]{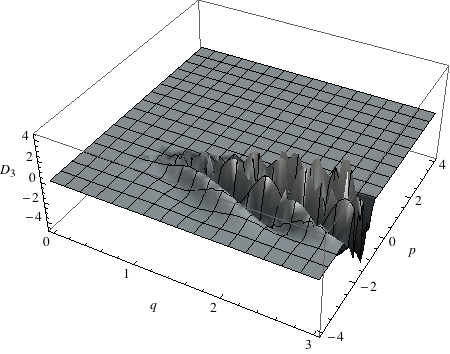}
\includegraphics[width=.3\textwidth]{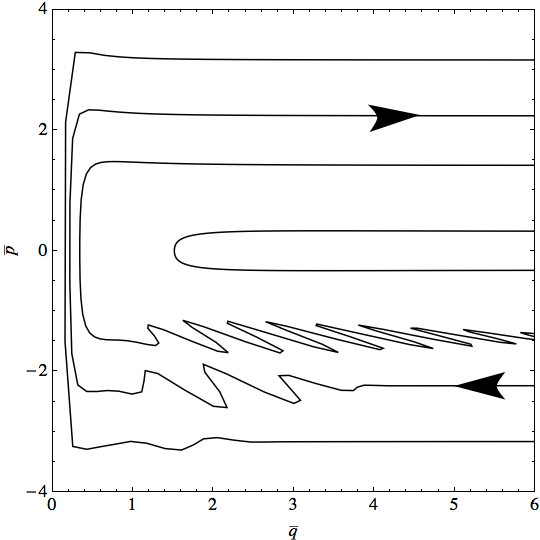}
\includegraphics[width=.3\textwidth]{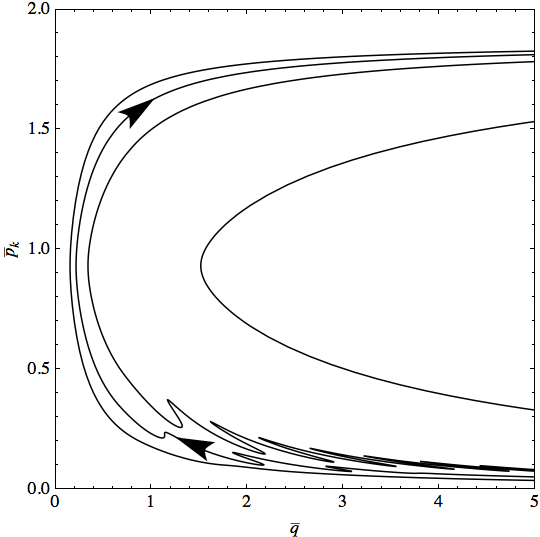}
\caption{\small From left to right: (i) plot of the delay function $D_3$ which defines the transformation between the reduced phase space used to derive the semiclassical dynamics and the reduced phase space used to describe it; (ii) the semiclassical evolution in the $(\bar{q},\bar{p})$-plane for $k=1$, $c_w=1$, $\check{\mathrm{H}}_{true}=\frac{1}{240},\frac{1}{12},\frac{5}{24}, \frac{5}{12}$; (iii) the semiclassical evolution in the $(\bar{q},\bar{p}_k)$-plane.} 
\label{d3}
\end{figure}

\begin{figure}[t]
\centering
\includegraphics[width=.38\textwidth]{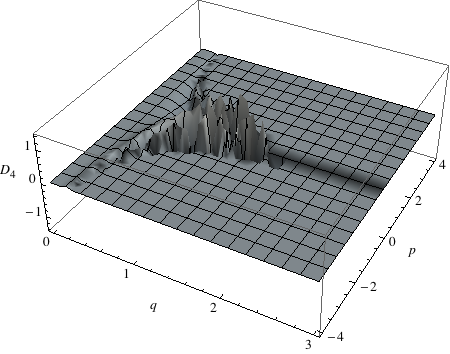}
\includegraphics[width=.3\textwidth]{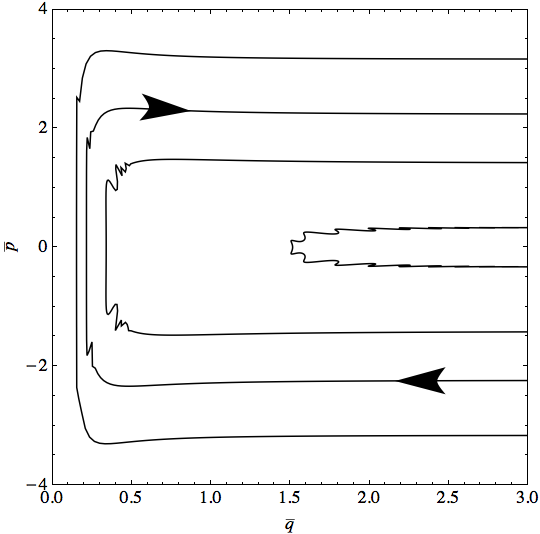}
\includegraphics[width=.3\textwidth]{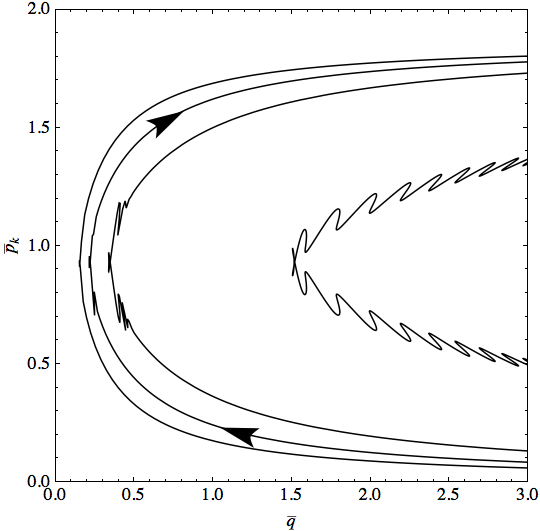}
\caption{\small From left to right: (i) plot of the delay function $D_4$ which defines the transformation between the reduced phase space used to derive the semiclassical dynamics and the reduced phase space used to describe it; (ii) the semiclassical evolution in the $(\bar{q},\bar{p})$-plane for $k=1$, $c_w=1$, $\check{\mathrm{H}}_{true}=\frac{1}{240},\frac{1}{12},\frac{5}{24}, \frac{5}{12}$; (iii) the semiclassical evolution in the $(\bar{q},\bar{p}_k)$-plane.} 
\label{d4}
\end{figure}

The first example is
\begin{align}D_1(q,p):=40\cdot \eta(\frac{q}{p+\frac{k}{q}})\eta(\frac{-pq-10k}{p^2-\frac{k^2}{q^2}})\eta(-p^2+\frac{k^2}{q^2}).\end{align}
The figure \ref{d1} includes a 3-dimensional plot of the delay function $D_1$ and the semiclassical phase space portrait in the planes $(\bar{q},\bar{p})$ and $(\bar{q},\bar{p}_k)$ mapped from the planes $(q,p)$ and $(q,p_k)$, where the semiclassical motion was derived. The clock transformation $T\mapsto \bar{T}=T+D_1(q,p)$ occurs only in the phase space region where the classical universe contracts. As a result, the semiclassical motion in this region deviates from the one derived with the reference clock $\bar{T}$. Moreover, the dynamics of the variable $\bar{p}_k$ that is coupled to $\bar{q}$ and $\bar{p}$ deviates during the same phase of evolution, i.e. for small and contracting $\bar{q}$. Finally, we observe that away from the bounce the dynamics coincides with the dynamics derived in $\bar{T}$ and represented in the figure \ref{figure2}.

The second example is
\begin{align}D_2(q,p):=q\cdot\eta(p^2-\frac{k^2}{q^2}).\end{align}
The figure \ref{d2} includes the plot of $D_2$ and the portrait of the quantum motion in the reduced phase space $(\bar{q},\bar{p},\bar{p}_k)$ that is different by $D_2$ from the reduced phase space $(q,p,p_k)$ in which the quantum motion is derived. The respective clock transformation occurs only in the classically forbidden region and as a result, the semiclassical dynamics exhibits some deviations from the dynamics of the figure \ref{figure2} precisely in this region. Finally, we observe that away from the bounce the dynamics coincides with the dynamics of the figure \ref{figure2}.

The next example is 
\begin{align}D_3(q,p):=20 \cdot\sin(4\pi \frac{pq}{p^2-\frac{k^2}{q^2}}) \eta(\frac{q}{p+\frac{k}{q}})\eta(\frac{-pq-10k}{p^2-\frac{k^2}{q^2}})\eta(-p^2+\frac{k^2}{q^2}).\end{align}
The figure \ref{d3} includes the plot of $D_3$ and the portrait of the quantum motion in the reduced phase space $(\bar{q},\bar{p},\bar{p}_k)$ that is different by $D_3$ from the reduced phase space $(q,p,p_k)$ in which the quantum motion is derived. The respective clock transformation occurs only in the phase space region where the classical universe contracts and it is oscillatory. As a result, the semiclassical dynamics exhibits some oscillatory deviations from the dynamics of the figure \ref{figure2} precisely in this region. Finally, we observe that away from the bounce the dynamics coincides with the dynamics of the figure \ref{figure2}.

The final example is 
\begin{align}D_4(q,p):=1.5\cdot \sin(4\pi \frac{pq}{p^2-\frac{k^2}{q^2}}) q^2 \eta(3p^2-\frac{3k^2}{q^2}).\end{align}
The figure \ref{d4} includes the plot of $D_4$ and the portrait of the quantum motion in the reduced phase space $(\bar{q},\bar{p},\bar{p}_k)$ that is different by $D_4$ from the reduced phase space $(q,p,p_k)$ in which the quantum motion is derived. The respective clock transformation occurs only in the classically forbidden region and is oscillatory. As a result, the semiclassical dynamics exhibits some oscillatory deviations from the dynamics of the figure \ref{figure2} precisely in this region. Finally, we observe that away from the bounce the dynamics coincides with the dynamics of the figure \ref{figure2}.

Let us summarise what we have learnt from the above numerical examples. The choice of clock and associated reduced phase space determines the quantum dynamics obtained by subsequent quantisation. The method of phase space portraits based on coherent states reveals that the dissimilarities in quantum dynamics are non-trivial and can be very large. However, as the studied examples suggest, away from where the quantum effects dominate, all quantum dynamics admit a fixed classical limit irrespectively of the employed internal clock. As we show next, this is in fact a universal property.

\section{Generalisation of the result}

In the previous section we showed that semiclassical dynamics of the Bianchi type I model based on different internal clocks are different. This is visible in the vicinity of the bounce, where quantum corrections dominate. Away from the bounce semiclassical evolutions coincide irrespectively of the employed clock. This is a non-trivial result since the mere vanishing of semiclassical corrections does not imply that the dynamics convergences to the same classical solution. In what follows we show how this result can be generalised to all models defined in phase spaces of any dimension. We focus on the semiclassical dynamics and use phase space portraits. We assume that the quantised models satisfy the classical limit in the sense that their semiclassical dynamics converges in the asymptotic future and past to the classical dynamics (without specifying to which classical solution exactly). We start with a useful definition and then consider two cases, a restricted and a general one.

{\it Definition.} Suppose there is a bounded region in the phase space, ${\it R}$, such that the semiclassical trajectories of a given model cross it only for a finite amount of time and move away from it both in the asymptotic past and future. For instance, in cosmological bouncing models it is a region where the volume of the universe is small and the expansion rate large. At the fully quantum level this bounded region could mean some subset of the Hilbert space, through which the evolution of any initial state may take place, but which does not contain the asymptotic states.


\begin{figure}[t]
\centering
\begin{subfigure}[t]{0.45\textwidth}\begin{center}
\includegraphics[scale=0.34]{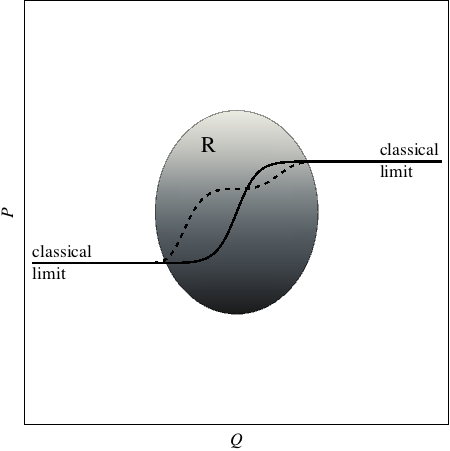}
\caption{\small Schematic representation of a pseudo-canonical transformation confined to a bounded ${\it R}$. The solid curve represents a semiclassical trajectory which is transformed with a pseudo-canonical transformation into another semiclassical trajectory represented with a dashed curve. Outside the bounded region ${\it R}$, the transformation is trivial and both curves coincide.} 
\label{GAA}\end{center}
\end{subfigure}
\hspace{0.05\textwidth}
\begin{subfigure}[t]{0.45\textwidth}\begin{center}
\includegraphics[scale=0.34]{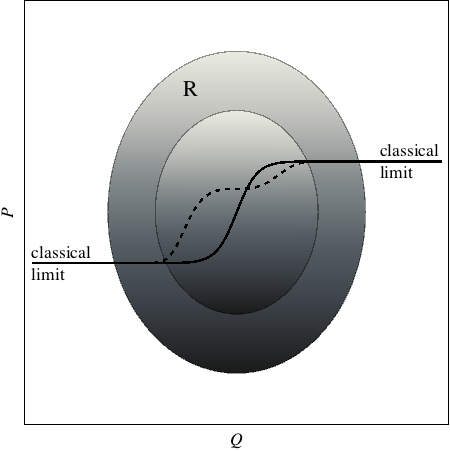}
\caption{\small Schematic representation of a pseudo-canonical transformation confined to ${\it R}$ that is enlarged with respect to ${\it R}$ of the left panel. The solid curve represents a semiclassical trajectory which is transformed with a pseudo-canonical transformation into another semiclassical trajectory represented with a dashed curve. Despite of the enlarged ${\it R}$, where the pseudo-canonical transformation is not trivial, the difference between the two curves remains unchanged. The semiclassical trajectory reaches a classical limit already inside the enlarged ${\it R}$ and hence, the pseudo-canonical transformation does not change it.} 
\label{GAB}\end{center}
\end{subfigure}
\end{figure}


{\it Restricted case.} Let us first consider a pseudo-canonical transformation (\ref{cons}) which is trivial outside a bounded region ${\it R}$, based on a delay function $D$ which vanishes outside ${\it R}$. This type of clock transformation is represented schematically in Fig. \ref{GAA}. As discussed in Sec. \ref{qc}, classical Hamiltonians $H(q^I,p_I)$ and $H(\bar{q}^I,\bar{p}_I)$ which act in two different reduced phase spaces related by a special pseudo-canonical transformation are assigned the same quantum operator in the Hilbert space $\mathcal{H}$. Subsequently, as discussed in Sec. \ref{spsp}, they are given formally the same semiclassical approximation in terms of $\check{H}(q^I,p_I)$ and $\check{H}(\bar{q}^I,\bar{p}_I)$ that generate formally the same semiclassical dynamics in the respective reduced phase spaces. In other words, the semiclassical solution in one reduced phase space is deduced from the semiclassical solution in the other one by formal replacement of $(q^I,p_I,t)$ by $(\bar{q}^I,\bar{p}_I,\bar{t})$. Since outside ${\it R}$ the pseudo-canonical transformation is trivial (i.e., the identity), the variables $(q^I,p_I)$ and $(\bar{q}^I,\bar{p}_I)$ become physically identical there. This shows that the formally identical semiclassical dynamics inside ${\it R}$ become physically identical  outside ${\it R}$. We conclude that a given asymptotical classical solution in remote past is evolved to a fixed asymptotical classical solution in remote future irrespectively of the pseudo-classical transformation in the bounded region ${\it R}$.

{\it General case.}  We assumed that asymptoticly the semiclassical dynamics for any clock becomes classical. The latter means that $\check{H}(q^I,p_I)\rightarrow {H}(q^I,p_I)$ and $\check{H}(\bar{q}^I,\bar{p}_I)\rightarrow {H}(\bar{q}^I,\bar{p}_I)$ along any trajectory in the limit $t\rightarrow\pm\infty$ and $\bar{t}\rightarrow\pm\infty$. We recall that by construction (see Sec. \ref{gentheory}), the classical evolution is invariant with respect to pseudo-canonical transformations. Suppose we enlarge the bounded region ${\it R}$ (where a pseudo-canonical transformation is non-trivial) continuously and indefinitely along semiclassical trajectories. Since the  semiclassical trajectories converge to classical ones and since the classical trajectories are invariant with respect to pseudo-canonical transformations, asymptotically the semiclassical trajectories must be invariant too. We conclude that asymptotically the phase space portraits become physically identical irrespectively of any pseudo-canonical transformation (see Fig. \ref{GAB}).

{\it Bianchi-I model.} For the semiclassical Bianchi type I model, the asymptotic vanishing of the physical differences between the semiclassical trajectories in $(q,p_k)$ and $(\bar{q},\bar{p}_k)$ follows from a simple fact that the semiclassical Hamiltonian of (\ref{semiH}) becomes classical in the limit of large volume (or, large $q$), i.e. for ${q\rightarrow \infty}$ we have $\check{\mathrm{H}}_{true}(q,p,k)\rightarrow {\mathrm{H}}_{true}(q,p,k)$. We notice that the semiclassical Hamiltonian of the flat Friedmann model studied in \cite{2M15} satisfies an analogous limit. However,  the studied phase space of \cite{2M15} is two-dimensional and the conservation of energy is sufficient to impose the invariance of the asymptotic states. In the present case this is not sufficient because the values of a Hamiltonian in a higher dimensional phase space do not fix asymptotic trajectories unambiguously. The present result holds even if some classically conserved quantities are not conserved at the quantum level.

\section{Discussion}

The present paper concerns the question of the relation between quantum dynamics of relativistic models and the choice of internal clock. For this purpose we employed a singular cosmological Bianchi Type I model with a six dimensional reduced phase space. The quantum dynamics issued from the proposed quantisation replaced the classical singularity with a quantum bounce. The quantum dynamics was initially derived with a fluid variable playing the role of internal clock. Then we transformed the derived quantum dynamics into quantum dynamics with respect to other internal clocks by means of a previously developed theory of pseudo-canonical transformations. We made a choice of several internal clocks and compared the respective quantum dynamics by the phase space portrait method. The latter provides the semiclassical-level characterisation of quantum dynamics. In consistency with our previous result on a simpler Friedmann-Robertson-Walker model \cite{2M15}, we showed that many details of the quantum bounce strongly depend on the clock employed in quantisation. The new and vital question studied for the first time herein was whether the asymptotic classical solutions away from the bounce are joined by semiclassical bouncing dynamics in a fixed pattern regardless of the particular clock used for quantisation.

Our analysis confirms that the answer to the above question is affirmative. This finding is, to our best knowledge, the first demonstration of the predictive power of background-independent quantisations of gravity despite the fact that the choice of clock influences the quantum and semiclassical dynamics. We do not claim that our result solves the time problem in quantum gravity. Neither do we propose a satisfactory re-interpretation of quantum theory in the context of gravitational systems. However, we believe that we provide a detailed theory of an important and unavoidable aspect of quantum gravity models: the clock-dependence of the concept of dynamics. 

There may exist more clock-invariant features of quantum models of gravity but we do not know them at present. For the moment, our result suggests that the physical meaning in models of quantum gravity should be assigned exclusively to the asymptotic classical solutions and a causal relation between them. The latter is a purely quantum prediction. This approach resembles the scattering matrix formalism as the data is limited to the asymptotic states of a physical system. 

In view of the presented result one should ask why the usual formalism of quantum mechanics which relies on the concept of a fixed time and a unique quantum dynamics works well for laboratory systems. The answer seems to be intuitively clear, nevertheless we will investigate it in detail elsewhere. Quantum laboratory systems are parts of larger systems that include classical environments which in particular provide clocks. Suppose that we have obtained a set of coupled Hamilton and Schr\"odinger equations which describe the evolution of an environment (classical) and of a laboratory (quantum) system, respectively. Let us consider a pseudo-canonical transformation of this system of equations such that the transformation of the internal clock involves only variables of the environment whose evolution is described by the Hamilton equations. It is expected that such a transformation does not induce quantum effects in the evolution of the laboratory system and merely introduces a change in the units of time in the respective Schr\"odinger equation. Once the classical environments cease to exist, there are no longer privileged internal clocks and the usual formalism of quantum mechanics must break down.

\section{Acknowledgments} 

This work was supported by Narodowe Centrum Nauki by decision No. DEC-2013/09/D/ST2/03714.

\appendix
\section{Shape function}
\label{A}
The name of the {\it shape function} denoted by `$p_k$' is justified by the fact that it determines the relative values of the three scale factors. In terms of the canonical variables they read:
\begin{align}\nonumber
a_1&=q^{\frac{2}{c_w}}e^{-2p_k(\sqrt{3}\sin\alpha+\cos\alpha)+2\frac{p_{\alpha}}{k}(\sqrt{3}\cos\alpha+\sin\alpha)}\\\label{AA}
a_2&=q^{\frac{2}{c_w}}e^{2p_k(\sqrt{3}\sin\alpha-\cos\alpha)-2\frac{p_{\alpha}}{k}(\sqrt{3}\cos\alpha-\sin\alpha)}\\ \nonumber
a_3&=q^{\frac{2}{c_w}}e^{4(p_k\cos\alpha+\frac{p_{\alpha}}{k}(\sqrt{3}\sin\alpha)}\end{align}
From the eq. (\ref{AA}) it follows that the shape function describes the evolution of the relative sizes of the three spatial dimensions,
\begin{align}\left(\ln\frac{a_i}{a_j}\right)(T)=p_k(T) \cdot const.+const.\end{align}

\section{Delay functions in singular models}\label{C}
We assume that delay functions $D(q,p)$ vanish at the singularities where the classical trajectories terminate. Computations show that neglecting this assumption may spoil the singularity resolution. The assumption reads:
\begin{align}\lim_{T~\rightarrow~ T_{S}}D(q(T),p(T))=0,\end{align}
where $(q(T),p(T))$ is a classical trajectory terminating at $T_{S}$. In terms of new variables introduced in Eq. (\ref{coordtrans}) the above assumption takes the following form:
\begin{equation}
\forall~y>0~~\lim_{x\rightarrow -\frac{k}{y}^-} E(x,y)=0=\lim_{x\rightarrow \frac{k}{y}^+}  E(x,y),
\end{equation}
where $D(q,p)=E(x(q,p),y(q,p))$ and the ranges of $x$ and $y$ were given below Eq. (\ref{coordtrans}). 

Last technical remark concerns the relation (\ref{coordtrans}) which is ill-defined for $p^2=\frac{k^2}{q^2}$ or, equivalently for $y=0$. A careful analysis of the relation (\ref{coordtrans}) shows that in order to ensure the continuity of $E(x(q,p),y(q,p))$ in $(q,p)$ we must impose
\begin{align}
\lim_{y\rightarrow0^+} E(\frac{\sqrt{k^2+q_0^2y^2}}{y},y)=\lim_{y\rightarrow0^-} E(-\frac{\sqrt{k^2+q_0^2y^2}}{|y|},y)=D(q_0,\frac{k}{q_0}),\\
\lim_{y\rightarrow0^+} E(-\frac{\sqrt{k^2+q_0^2y^2}}{y},y)=\lim_{y\rightarrow0^-} E(\frac{\sqrt{k^2+q_0^2y^2}}{|y|},y)=D(q_0,-\frac{k}{q_0}),
\end{align}
where $q_0>0$. The delay functions studied as examples in Sec. (\ref{numex}) satisfy all the above conditions.

\section{Formulas for parameters of the semiclassical solution}\label{B}
\begin{align}\nonumber
A_0=\frac{\lambda_1\frac{k}{c_w}}{c_w\sqrt{\lambda_2(1+\lambda_1(\frac{k}{c_w})^2)}},~~\omega_0=2\sqrt{\frac{\lambda_2}{1+\lambda_1(\frac{k}{c_w})^2}}\\ 
A_1=\frac{2\frac{k}{c_w}(B\lambda_2-\lambda_1)}{c_w\sqrt{(1+\lambda_1(\frac{k}{c_w})^2+\lambda_2A-\lambda_2B(\frac{k}{c_w})^2)^2-4\lambda_2A}}\\\nonumber
\omega_{\pm}=\frac{\sqrt{2\lambda_2}2}{\sqrt{1+\lambda_1(\frac{k}{c_w})^2+\lambda_2A-\lambda_2B(\frac{k}{c_w})^2\pm\sqrt{(1+\lambda_1(\frac{k}{c_w})^2+\lambda_2A-\lambda_2B(\frac{k}{c_w})^2)^2-4\lambda_2A}}}
\end{align}

\section{Quantum Bianchi I Hamiltonian}\label{D}
The value of the integral (\ref{QHBI}) depends on the particular family of the affine coherent states (\ref{ACS}) employed in the quantisation map. We set the fiducial vector as
\begin{equation}\label{fiducial}
\psi^{\nu}(x)= \frac{e^{-\frac{ \nu}{4} \left(\frac{K_1(\nu)}{K_2(\nu)} x +\frac{K_2(\nu)}{K_1(\nu)x}\right)}}{\sqrt{2 x \, K_0(\nu)}},~ \, x>0,~ \nu>0,
\end{equation}
where $K_r(z)$ is a modified Bessel function. The parameter $\nu$ is free and enumerates a set of families of coherent states and, in result, a set of  quantisation maps. Now we compute (\ref{QHBI}):
\begin{align}\nonumber
&\langle x|\hat{\mathrm{H}}_{true}|x'\rangle =\frac{\delta(x-x')c_w^2}{24}\left(P^2+(\mathcal{J}-\frac{k^2}{c_w^2}\mathcal{I})Q^{-2}\right)\\
&~~-\frac{c_w}{12\pi}\int_0^{\infty}\frac{\ud q}{q}\left(\frac{k}{q}\frac{\cos\left(\frac{|k|}{qc_w}(x-x')\right)}{(x-x')^2}-\frac{\sin\left(\frac{|k|}{qc_w}(x-x')\right)}{(x-x')^3}\right)\bar{\psi^{\nu}}(x'/q)\psi^{\nu}(x/q),
\end{align}
where $Q=x$, $P=\frac{1}{i}\frac{\partial~}{\partial x}$, $\mathcal{J}=\frac{1}{4}\left(1+\nu\frac{K_0(\nu)}{K_1(\nu)}\right)$ and $\mathcal{I}=\left(\frac{K_2(\nu)}{K_1(\nu)}\right)^2$. This operator is nonlocal and very difficult to analyse. Therefore, we restrict our analysis to the phase space portrait of $\hat{\mathrm{H}}_{true}$.


\begin{thebibliography}{99}


\bibitem{Ku}
K. Kuchar, 
Time and interpretation of quantum gravity 
in {\it Proceedings of the 4th Canadian Conference on General Relativity and Relativistic Astrophysics}, World
Scientific, Singapore, (1992).

\bibitem{misner0}
C. W. Misner, Feynman Quantization of General Relativity, Rev. Mod. Phys., vol. 29, no.3, pp. 497--509 (1957)

\bibitem{1M15}
P. Ma\l kiewicz, Multiple choice of time in quantum cosmology,  Class. Quantum Grav. 32 (2015) 135004;


\bibitem{M12}
P. Ma\l kiewicz, Reduced phase space approach to Kasner universe and the problem of time in quantum theory, Class. Quantum Grav. 29 (2012) 075008.

\bibitem{BI} 
H. Bergeron, A. Dapor, J.-P. Gazeau, P. Ma\l kiewicz, Smooth Bounce in Affine Quantization of a Bianchi I,  Phys. Rev. D 91, 124002 (2015); 

\bibitem{2M15}
P. Ma\l kiewicz, What is Dynamics in Quantum Gravity?, arXiv:1505.04730.

\bibitem{adm}
R. Arnowitt S. Deser and C. W. Misner, The dynamics of General Relativity in {\it Gravitation: an introduction to current research} edited by Louis Witten (Wiley, 1962) chapter 7 pp 227--265

\bibitem{Sch}
B. F. Schutz, Hamiltonian Theory of a Relativistic Perfect Fluid, Phys. Rev. D 4 (1971) 3559



\bibitem{cwm3} C. W. Misner, Minisuperspace, in {\it Magic Without
Magic: John Archibald Wheeler}, p. 441, edited by J. R. Klauder (W.H.
Freeman and Company, San Francisco, 1972).

\bibitem{AbMa}
R. Abraham, J.E. Marsden, 
"Foundations of Mechanics", The Benjamin/Cummings Publishing Company, Inc. (1978).

\bibitem{ADM61}
R. Arnowitt S. Deser and C. W. Misner,
Heisenberg Representation in Classical General Relativity, Nuovo Cimento 19, 668 (1961)

\bibitem{JPt}
J.-P. Gazeau, Coherent States in Quantum Physics, Wiley-VCH, 2009

\bibitem{klauderscm}  J. R. Klauder, Enhanced Quantization: A Primer,
J. Phys. A: Math. Theor. \textbf{45} (2012) 285304--1--8 ; [arXiv:1204.2870];
Completing Canonical Quantization, and Its Role in
Nontrivial Scalar Field Quantization, [arXiv:1308.4658].


\end{thebibliography}
\end{document}